\RequirePackage{color}

\documentclass[11pt,a4paper]{article}
\pdfoutput=1
\usepackage{jheppub}

\usepackage{latexsym,amsmath,amsfonts,amssymb}
\usepackage{graphicx}
\usepackage{bbm}
\usepackage{multirow}
\usepackage{wrapfig}

\newcommand{\be}{\begin{equation}}
\newcommand{\ee}{\end{equation}}

\newcommand{\half}{\frac{1}{2}}

\newcommand{\barray}{\begin{eqnarray}}
\newcommand{\earray}{\end{eqnarray}}

\newcommand{\cA}{\mathcal{A}}

\newcommand{\cC}{\mathcal{C}}

\newcommand{\cF}{\mathcal{F}}

\newcommand{\cI}{\mathcal{I}}

\newcommand{\cL}{\mathcal{L}}

\newcommand{\cN}{\mathcal{N}}
\newcommand{\cO}{\mathcal{O}}

\newcommand{\cR}{\mathcal{R}}
\newcommand{\cS}{\mathcal{S}}

\newcommand{\beq}{\begin{equation}}
\newcommand{\eeq}{\end{equation}}

\DeclareMathOperator{\Tr}{Tr}

\def\drawbox#1#2{\hrule height#2pt 
        \hbox{\vrule width#2pt height#1pt \kern#1pt \vrule width#2pt}
              \hrule height#2pt}

\def\Fund#1#2{\vcenter{\vbox{\drawbox{#1}{#2}}}}
\def\Asym#1#2{\vcenter{\vbox{\drawbox{#1}{#2}
              \kern-#2pt       
              \drawbox{#1}{#2}}}}

\def\fund{\Fund{6.5}{0.4}}
\def\afund{\overline \fund}

\title{Theories of Class $\cS$ and New $\cN=1$ SCFTs}

\author{James McGrane and Brian Wecht}
  
\affiliation{Centre for Research in String Theory\\
Queen Mary University of London\\
London E1 4NS\\
United Kingdom}

\abstract{We describe an infinite two-parameter subfamily of theories of class ${\cal S}$ where dialing one of the parameters interpolates between Gaiotto's $T_N$ theory and a theory of $N^2$ free hypermultiplets. After using the reduced superconformal index to study the operator content, we use these theories to construct new $\cN=1$ SCFTs and then examine the flows between them.}

\preprint{QMUL-PH-14-22}

\begin{document}

\maketitle

\section{Introduction}
\label{sec:intro}

We are living in a golden age of quantum field theories. The diversity of theories available to study is astonishing, and due to the technological advances of recent years, many strongly coupled theories that had been considered intractable are now able to be investigated. There are no better examples of this than the supersymmetric compactifications of the still-mysterious six-dimensional $(2,0)$ theory. These exotic theories, which generically do not have free-field limits, are nevertheless rather understandable, and many quantities of interest (e.g.~operator dimensions) are calculable. Although a great deal of progress has been made on compactifications of the $(2,0)$ theory to two and three dimensions, in the present work we will be most interested in the four-dimensional theories that come from compactifying on a (punctured) Riemann surface. This compactification can be done in such a way as to preserve $\cN =2$ SUSY in four dimensions \cite{Maldacena:2000mw}, and the resulting theories are called theories of class $\cS$. 

For theories of class $\cS$, we need only specify two pieces of compactification data in order to determine the theory: the genus $g$ of the Riemann surface, and the pole structures of the punctures. Since any (punctured) Riemann surface with genus $g>1$ can be described with a suitable gluing of thrice-punctured spheres, we can describe any such theory by a set of these spheres, (called ``fixtures'' in \cite{Chacaltana:2010ks}) with some subset of punctures connected by cylinders, so that the final object has the required genus and punctures. The most famous examples of fixtures, and the flagship examples for novel four-dimensional SCFTs, are Gaiotto's $T_N$ theories \cite{Gaiotto:2009we}. Using this construction, the punctures correspond to global symmetries, and the cylinders correspond to gauge symmetries. In this manner, we can construct infinitely many $\cN=2$ SCFTs.

Because of the tremendous amount of freedom available to us in constructing such theories, the landscape of theories of class $\cS$ still seems like the Wild West, and although general principles for these theories are known, not overly many specific examples have been explored. Much as in the case with D3-branes at the tip of a toric singularity, it would be useful to have a nice infinite family of theories to play with, like the $Y_{p,q}$ or $L_{p,q,r}$ theories. In this paper, we point out the existence of such an infinite family, which includes and generalizes Gaiotto's $T_N$ theories. For reasons that will become apparent in the body of the paper, we refer to these as the $T_{N,k}$ theories.

Many properties of these theories are still mysterious. For theories of class $\cS$, although much is known about the Coulomb branch via the Seiberg-Witten curve, the Higgs branch remains relatively unexplored. One reason is because, unlike in theories with Lagrangian descriptions, there is no candidate basis of UV-free fields one could use to build a list of Higgs branch operators. Thus, it remains unclear how to even find the Higgs branch operators, much less the intricate relationships between them.

One window we do have into the Higgs branch is through the superconformal index (SCI) \cite{Romelsberger:2005eg, Kinney:2005ej}, which is a useful tool for finding operators. In theories of class $\cS$, a reduced version of the SCI was found in \cite{Gadde:2011ik}, and it is possible to use this to infer the existence of some Higgs branch operators which are difficult to see from duality alone, along with some of their quantum numbers. 

Although it is far from obvious from the fixtures-and-punctures approach, we can similarly construct a huge variety of new $\cN=1$ theories. Geometrically, one way of doing this is to change the embedding of the Riemann surface in the 11-dimensional space by suitably twisting the normal bundle. The existence of certain $\cN=1$ supergravity solutions was first shown in \cite{Maldacena:2000mw}; these solutions were then shown to be part of a much larger set of solutions in \cite{Bah:2012dg}, and further supergravity solutions were found in \cite{Bah:2013qya}. Alternately, one could use a recently-discovered class of punctures \cite{Xie:2013gma,Xie:2013rsa,Bah:2013aha, Bonelli:2013pva} which preserve only $\cN=1$ SUSY. From a field theory perspective, although certain of these solutions arise at the endpoints of flows from theories of class $\cS$ \cite{Benini:2009mz}, the overwhelming majority are not known to do so. 

Another goal of the present work is to further the study of $\cN=1$ theories built out of class $\cS$ fixtures, as begun in \cite{Benini:2009mz} and continued in \cite{Bah:2012dg,Xie:2013gma,Xie:2013rsa,Bah:2013aha,Bah:2011je,Bah:2011vv,Maruyoshi:2013hja,Agarwal:2013uga,Tachikawa:2011ea, Agarwal:2014rua,Giacomelli:2014rna}. The study of these $\cN=1$ theories is still in its infancy, and many of their properties are unknown. In particular, it is not in general known which such theories are superconformal, and just as in conventional gauge theories, finding the IR phase of a given theory is often a difficult process. In \cite{Bah:2011je}, several such theories were analyzed, and flows between them were used to establish evidence for the existence or non-existence of the conformal fixed points. In the present work we re-examine these flows, and find a subtlety in the previous analysis which indicates that some of the theories not previously believed to flow to interacting conformal points may in fact do so.

The remainder of the paper is structured as follows. In Section \ref{sec:review}, we review some basic class $\cS$ technology. In Section \ref{sec:family}, we introduce a particularly interesting subfamily of theories of class $\cS$, the $T_{N,k}$'s, and describe some of their properties. In Section \ref{sec:SCI}, we review the superconformal index, and use it to elucidate further properties of the $T_{N,k}$ theories. In Section \ref{sec:flows}, we construct $\cN=1$ theories from the $T_{N,k}$'s, and describe some flows between them. Finally, in Section \ref{sec:wopunc}, we describe some initial attempts to construct theories in the manner of \cite{Bah:2011vv}. Various results are collected in appendices.

\section{Review}
\label{sec:review}

Even though many of the results in this work will be for $\cN=1$ theories, we will need to begin by reviewing some relevant $\cN=2$ technology. This will allow us to construct an interesting subclass of theories, which we will then explore in the remainder of the paper.

\subsection{Theories of Class \texorpdfstring{$\cS$}{S}}
\label{subsec:distler}

We begin with a brief review of theories of class $\cS$. This subsection roughly follows the format of \cite{Chacaltana:2010ks}. These theories are obtained by compactifying the six-dimensional $(2,0)$ theory on a Riemann surface $\cC$ with punctures. In this work we only consider theories coming from type $A_{N-1}$ six-dimensional $(2,0)$ theories; these theories arise on the worldvolume of a stack of $N$ M5-branes. 

In \cite{Gaiotto:2009we}, Gaiotto showed that the space of marginal couplings of these theories could be identified with the  moduli space of a curve $\cC_{g,h}$ with genus $g$ and $h$ punctures. Since then, these theories have seen a great deal of study, and it has been observed that the parameters defining the four-dimensional theory are completely determined by the two-dimensional compactification surface.  These defining parameters of the theory are, in addition to the genus $g$ of $\cC_{g,h}$, the location and type of the punctures on the surface. This data is encoded in the Seiberg-Witten curve, which is of the form $\lambda^N = \sum_{k=2}^{N}\lambda^{N-k}\phi_k$, where $\lambda$ is the Seiberg-Witten differential and $\phi_k$ are $k$-differentials ($k=2,...,N$). The $\phi_k$ will, in general, have poles at each of the punctures. Each puncture then can be characterized by its pole structure $\{p_k\} = \{p_2,p_3,...,p_n\}$ where $p_k$ is the order of the pole that $\phi_k$ has at the puncture. Then, for a given surface $\cC_{g,h}$ we can specify both the number of punctures as well as their individual pole structures.

\begin{wrapfigure}{r}{0.35\textwidth}
\begin{center}
\includegraphics[scale=0.15]{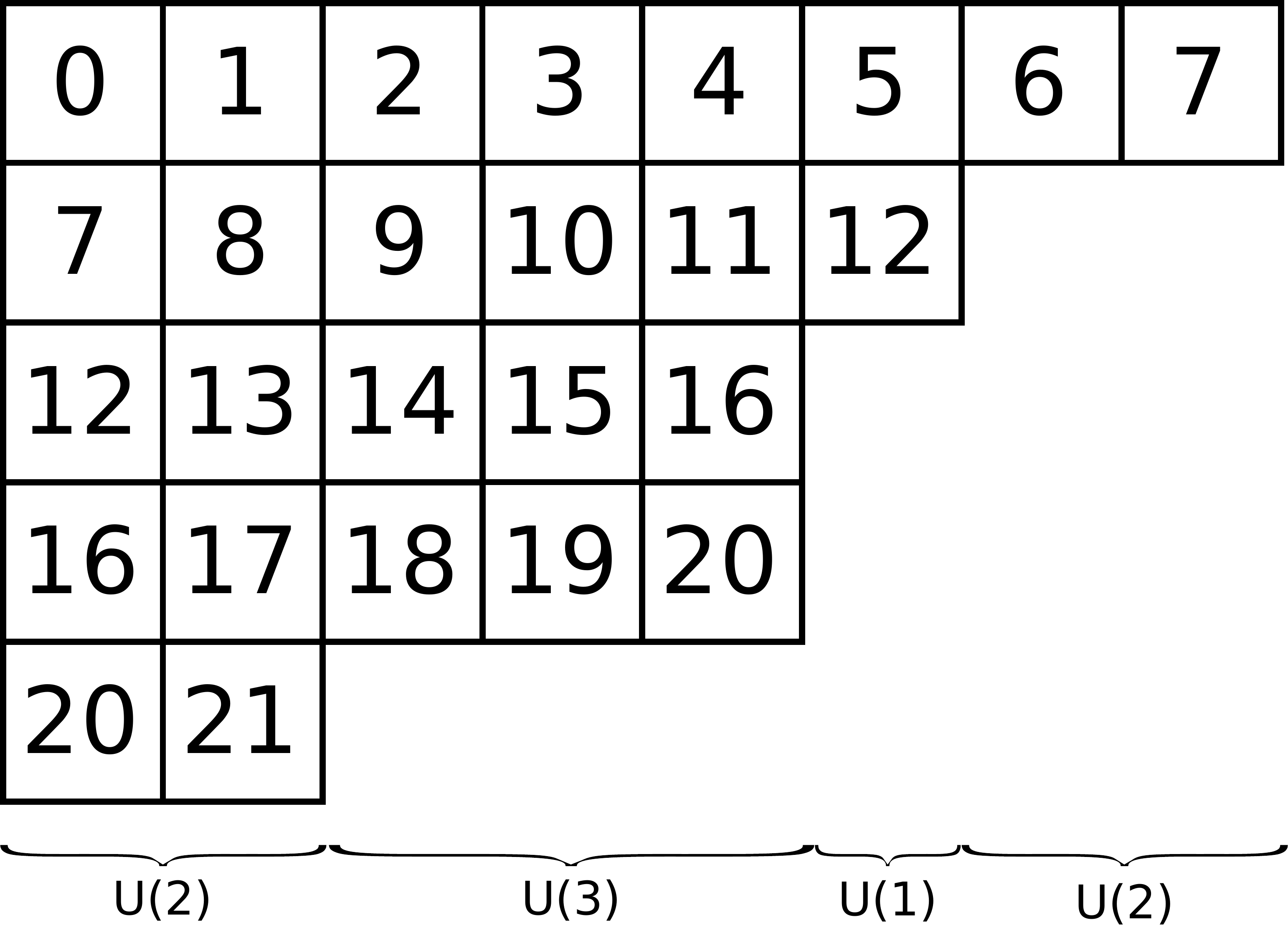}
\caption{A Young tableau for a regular puncture; the flavor symmetry associated to the puncture is $S\left(U(3) \times U(2)^2 \times U(1) \right)$.}
\label{fig:yt}
\end{center}
\end{wrapfigure}

Punctures come in two varieties, regular and irregular; which category a given puncture is in is determined by its pole structure. A \textit{regular} puncture is a puncture to which we can assign a Young tableau using the following rules\footnote{More generally for theories of class $\cS$, regular punctures are classified by embeddings of $SU(2)$ in the ADE Lie algebra of the six-dimensional theory. For class $\cS$ theories of type $A$ we can use Young tableaux, however, for type $A$ theories in the presence of an outer automorphism twist, or type $D$ or $E$ theories, this will not suffice. For more information see \cite{Chacaltana:2011ze, Chacaltana:2012ch, Chacaltana:2012zy, Chacaltana:2013oka, Chacaltana:2014jba, Chacaltana:2015bna}.}
\begin{itemize}
\item Draw a Young tableau with two boxes in a row;
\item For each $k=3,...,N$, if $p_k=p_{k-1}+1$ add a box to the current row, and if $p_k=p_{k-1}$ start a new row with one box.
\end{itemize}
All regular punctures must have $p_2=1$. As an example of a regular puncture, we have drawn the Young tableau in figure \ref{fig:yt}, for the puncture with pole structure $\{p_k\}=\{1,2,3,4,5,6,7,7,8,9,10,11,12,12,13,14,15,16,16,17,18,19,20,20,21\}$. The associated flavor symmetry is $SU(3) \times SU(2)^2 \times U(1)^3$.

The non-R global (flavor) symmetry group associated to a puncture is given by $G=S\left(\prod_{h}U(n_h)\right)$, where the product is over column heights of the Young tableau and $n_h$ is the number of columns with height $h$. The $S(...)$ means an overall $U(1)$ is removed. There are two special regular punctures worth highlighting. The first is a maximal puncture which has pole structure $\{1,2,...,N-1\}$ and flavor symmetry $SU(N)$; the corresponding Young tableau has one row of $N$ boxes. The second is a minimal puncture, which has pole structure $\{1,1,...,1\}$ and flavor symmetry $U(1)$; the corresponding Young tableau has one row of $2$ boxes and $N-1$ rows of 1 box each. 

\textit{Irregular}\footnote{``Irregular'' here is used in the sense of \cite{Chacaltana:2010ks}, and not in the same sense as most of the Hitchin system literature, e.g. \cite{Nanopoulos:2010zb}.} punctures are those punctures which do not satisfy the conditions for regular punctures, but do satisfy a different set of conditions whose structure we do not detail here; for useful discussions on irregular punctures see \cite{Chacaltana:2012ch} or \cite{Gaiotto:2011xs}.

\subsubsection{Fixtures and Cylinders}
\label{subsubsec:fixtures}
A {\it fixture} is a thrice-punctured sphere specified by the pole structure of each of the punctures. 
The quantity
\begin{equation}
d_k=1-2k+\left(\sum_{i=1}^3p_k^{(i)}\right),
\label{eq:dk}
\end{equation}
where the sum is over the punctures, gives us the number of Coulomb branch operators of dimension $k$. We can thus find the dimension of the Coulomb branch by summing over $k=2,...,N$. If the dimension of the Coulomb branch is zero then the fixture corresponds to a set of free hypermultiplets, and if the dimension of the Coulomb branch is greater than zero, then the fixture corresponds to a ``non-Lagrangian" SCFT\footnote{As usual, the phrase ``non-Lagrangian" merely means that no free-field UV description is known to exist, and not that such a description has been conclusively ruled out.}, or a combination of a non-Lagrangian SCFT and free hypers. Although the flavor symmetry of a fixture is usually just the product of the flavor symmetries associated to each of the punctures, there are some cases where the symmetry enhances.

One example of a fixture is one with two maximal punctures and one minimal puncture. This fixture has $d_k=0$ for all $k=2,...,N$ and corresponds to a theory of $N^2$ free hypermultiplets. A second useful example is the fixture with three maximal punctures. This fixture corresponds to the $T_N$ theory \cite{Gaiotto:2009we}. The $T_N$ has flavor symmetry $SU(N)^3$, when $N > 3$. The case $N=3$ is the $E_6$ SCFT of \cite{Minahan:1996fg}, and $N=2$ is a theory of 4 free hypermultiplets. The graded dimension of the Coulomb branch for the $T_N$ is $\{d_k\}=\{0,1,2,3,...,N-2\}$.

Punctures can be connected via {\it cylinders}, which correspond to a gauge group $G$ which must be a subgroup of the flavor symmetry group associated to each of the two punctures that it is connecting; this corresponds to gauging a flavor symmetry. As the cylinders get longer, the corresponding gauge coupling becomes weaker. Even for class $\cS$ theories of the same type, not every pair of punctures admits a cylinder connecting them; for the complete rules for type $A$ theories, see \cite{Chacaltana:2010ks}.

\subsubsection{S-Duality}
\label{subsubsec:sdual}

\begin{figure}[ht]
\begin{center}
\includegraphics[scale=0.1]{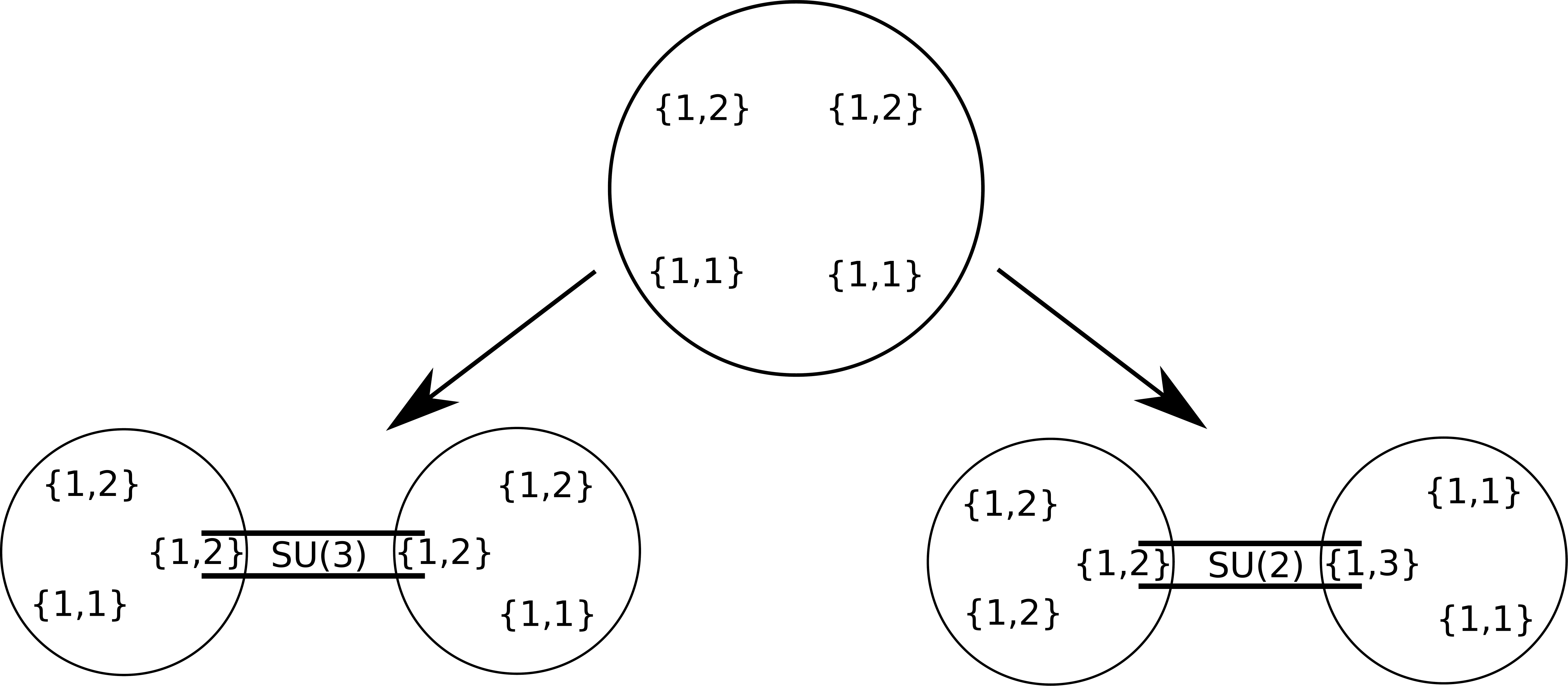}
\caption{The two degeneration limits of a punctured sphere with two maximal punctures and two minimal punctures. The picture on the left corresponds to the degeneration limit corresponding to an $SU(3)$ gauge theory with 6 fundamental hypermultiplets. The picture on the right corresponds to an $SU(2)$ gauge theory with one fundamental hypermultiplet and the $E_6$ SCFT, where an $SU(2)$ subgroup of the $E_6$ flavor symmetry is gauged.}
\label{fig:asdual}
\end{center}
\end{figure}

From the perspective of punctured surfaces, S-duality  corresponds to  different degeneration limits into thrice-punctured spheres connected by cylinders. As an example we look at the case of Argyres-Seiberg duality \cite{Argyres:2007cn}, which is depicted in Figure \ref{fig:asdual}.
The theory in question is derived by wrapping the six-dimensional $(2,0)$ $A_2$ theory on a Riemann surface with two maximal and two minimal punctures.

This theory can be decomposed into thrice-punctured spheres connected by cylinders in two ways. In one limit, there is an $SU(3)$ gauge theory with six hypermultiplets. In the other limit, there is an $SU(2)$ gauge theory with one hypermultiplet, where the $SU(2)$ gauges part of the global symmetry of  the $E_6$ SCFT of \cite{Minahan:1996fg}. Dualities of this form obey a set of consistency checks that were set out in \cite{Argyres:2007tq}.

\begin{figure}[ht]
\begin{center}
\includegraphics[scale=0.1]{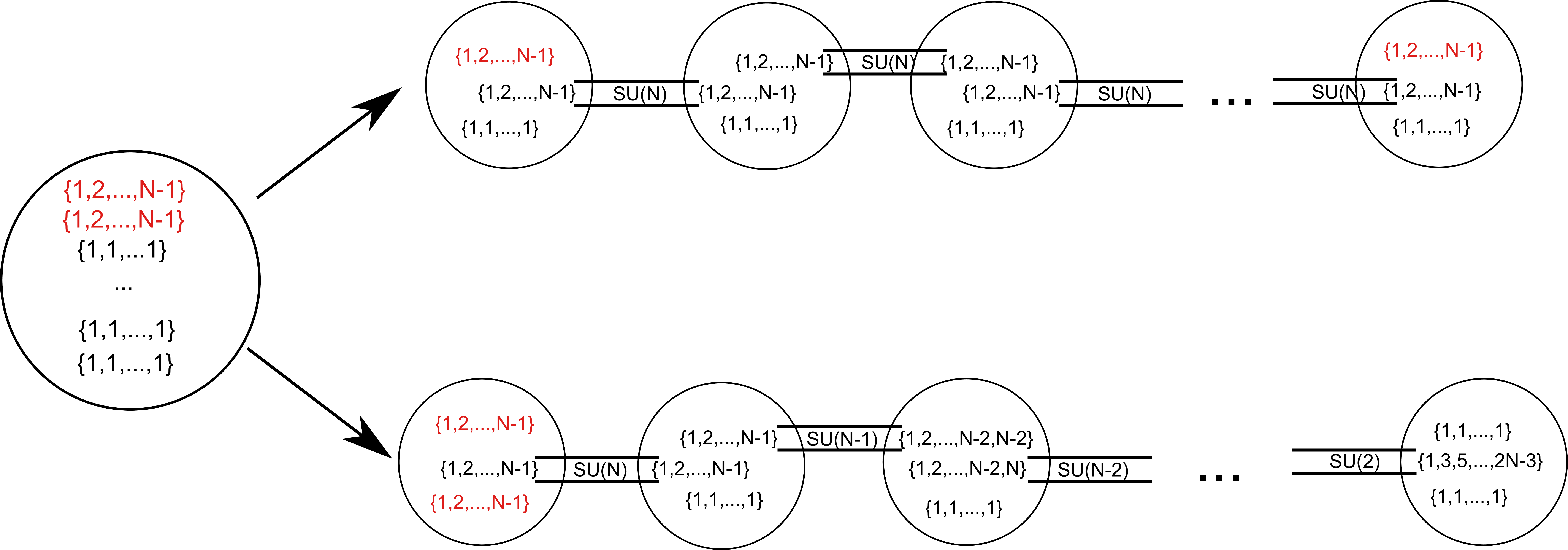}
\caption{Two different degeneration limits of a Riemann surface with two maximal punctures and $N-1$ minimal punctures. On the top is an $SU(N)^{N-2}$ gauge theory with bifundamental hypermultiplets. Each fixture by itself corresponds to $N^2$ free hypermultiplets and each cylinder corresponds to an $SU(N)$ gauge group which weakly gauges the flavor symmetries of the hypermultiplets. On the bottom is the $T_N$ coupled to a superconformal tail.}
\label{fig:gaiottodual}
\end{center}
\end{figure}

Gaiotto duality is another example of S-duality and relates an $SU(N)^{N-2}$ gauge theory to a non-Lagrangian theory, the $T_N$, coupled to a Lagrangian ``superconformal tail'' (more about this in the next section). This duality can be seen as two ways in which a genus 0 curve with two maximal punctures and $N-1$ minimal punctures can degenerate. These two ways are shown in figure \ref{fig:gaiottodual}.

\section{The \texorpdfstring{$T_{N,k}$}{TNk} Theories}
\label{sec:family}

In \cite{Gaiotto:2009we}, evidence was given for the existence of a one-parameter family of $\cN=2$ SCFTs, the $T_N$ theories. These theories are the low-energy energy limit of a stack of $N$ M5 branes wrapping a sphere with three maximal punctures. In the present work, we consider a related class of theories which will display a variety of interesting properties. These theories, which we will call $T_{N,k}$, come from $N$ M5-branes wrapping a sphere with two maximal punctures and a third puncture with pole structure $\{1,2,3,...,k-1,k,k,...,k\}$. The flavor symmetry of this theory is then $SU(N)^2 \times SU(k) \times U(1)$.

\begin{figure}[ht]\begin{center}
\includegraphics[scale=.15]{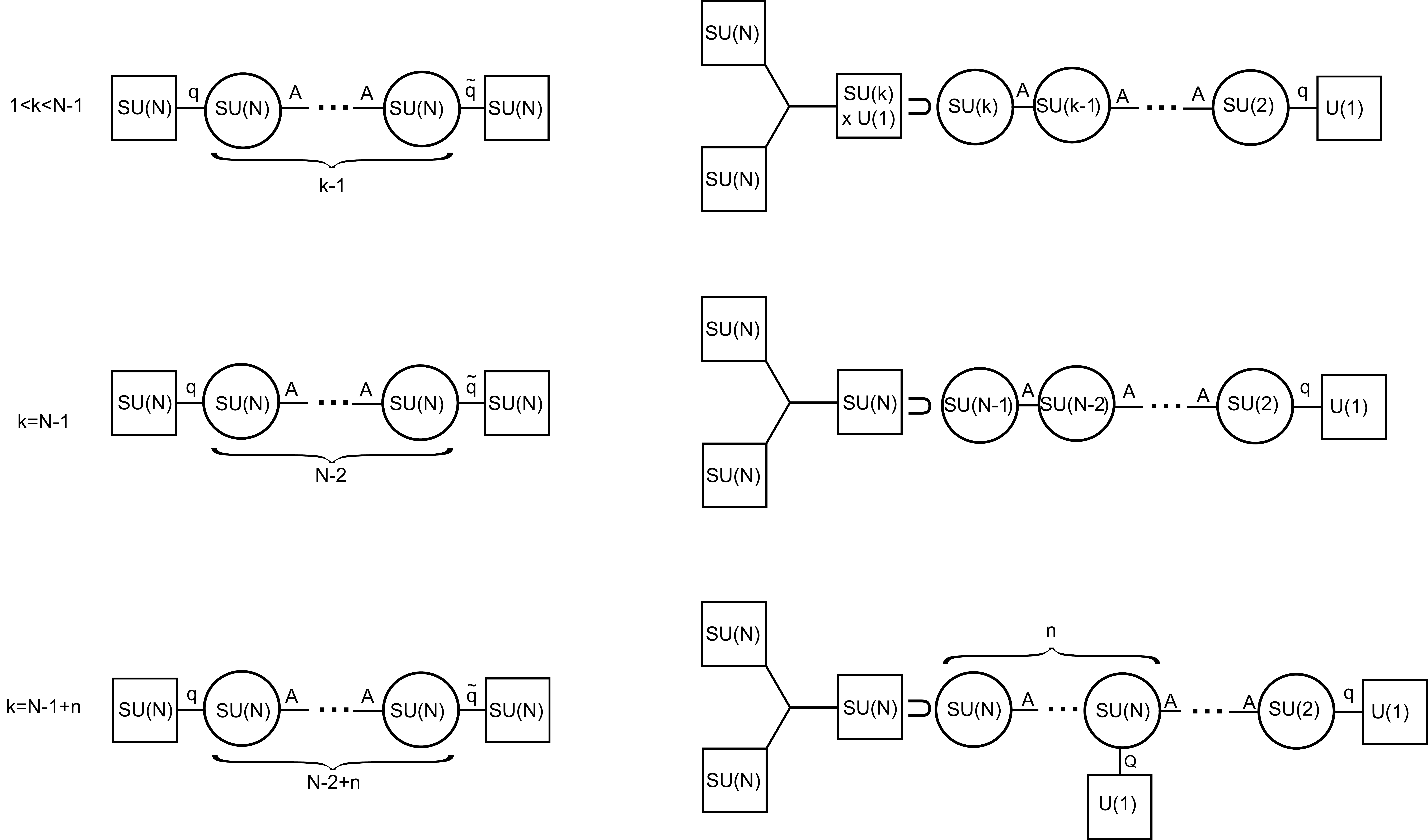}
\caption{Duality between $\cN=2$ linear quiver gauge theories (left) and $T_{N,k}$ theories coupled to an $\cN=2$ superconformal tail (right). Circles represent gauge symmetries, boxes represent flavor symmetries, and lines represent bifundamental hypers. Trivalent vertices represent $T_{N,k}$ theories. 
$\supset$ represents gauging of a subgroup of a flavor symmetry. In one duality frame, we have an $SU(N)^{k-1}$ gauge theory with bifundamental hypermultiplets. In the other frame, we have a $T_{N,k}$ coupled to a quiver theory with gauge groups of decreasing rank. In the case of $k=1$, we have $N^2$ free hypermultiplets in both duality frames, and for the case of $k=N-1$ we have a $T_N$ coupled to a superconformal tail. For all $k>N-1$ we have a $T_N$ coupled to $SU(N)^{k-N+1} \times SU(N-1) \times SU(N-2) \times \cdots \times SU(2)$ gauge theory.}
\label{fig:duality}
\end{center}\end{figure}

These theories form part of an interesting S-duality which is shown in figure \ref{fig:duality}. This duality corresponds to different ways in which a curve with two maximal punctures and $k$ minimal punctures can degenerate into thrice-punctured spheres connected by cylinders. This set of S-dualities generalizes the Gaiotto duality found in \cite{Gaiotto:2009we}. Gaiotto duality (see the middle row of figure \ref{fig:duality}) relates an $SU(N)^{N-2}$ gauge theory with bifundamental hypermultiplets to a $T_N$ coupled to a superconformal tail (i.e. an $SU(N) \times SU(N-1) \times \cdots \times SU(2)$ gauge theory with bifundamental hypermultiplets). A related natural question to ask is what theory is S-dual to an $SU(N)^{k-1}$ linear quiver gauge theory for general $k$. For the case $k>N-1$ one can see that the dual theory is again a $T_N$ coupled to a Lagrangian theory. This time the Lagrangian part is an $SU(N)^{k-N+1} \times SU(N-1) \times \cdots \times SU(2)$ gauge theory, as in the bottom row of figure \ref{fig:duality}. However, for the case $k<N-1$, we find that the dual theory is a $T_{N,k}$ coupled to a superconformal tail, as in the top row of figure \ref{fig:duality}. 

\begin{figure}[ht]\begin{center}
\includegraphics[scale=0.07]{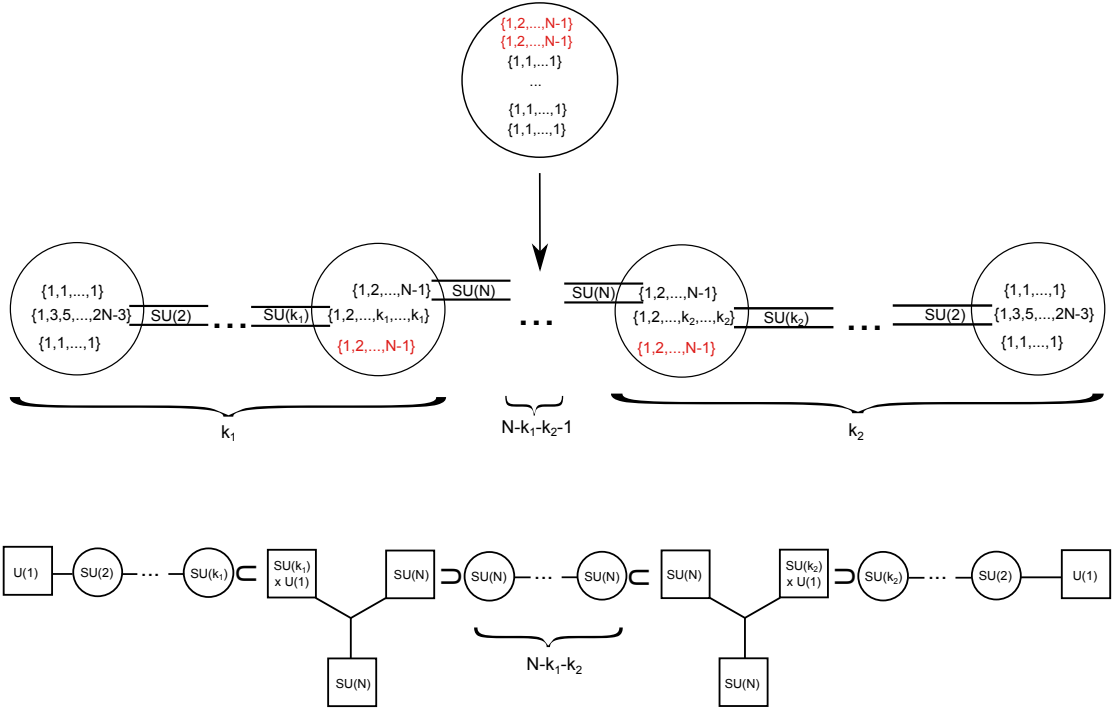}
\caption{Top: the degeneration limit of a surface with two maximal punctures and $N-1$ minimal punctures into thrice-punctured spheres connected by cylinders. The maximal punctures appear on the $k_1$-th sphere from the left and the $k_2$-th sphere from the right. Bottom: the quiver diagram for the corresponding theory, which contains a $T_{N,k_1}$ and a $T_{N,k_2}$. 
}
\label{fig:dualityb}
\end{center}\end{figure}

These theories also appear in another duality. In section \ref{subsubsec:sdual}, we reviewed how Gaiotto duality corresponds to two ways in which a genus 0 curve with two maximal punctures and $N-1$ minimal punctures can degenerate into thrice-punctured spheres connected by cylinders (see figure \ref{fig:gaiottodual}). The first way is to have one maximal puncture on each end sphere; this corresponds to the $SU(N)^{N-2}$ gauge theory. The second is to have the maximal punctures both on one end, which then corresponds to the $T_N$ coupled to the conformal tail. However, we can ask what happens when we degenerate the curve in such a way that the maximal punctures appear on fixtures in the middle of the curve rather than at the ends. In this case, the corresponding SCFT is a generalized quiver theory which involves $T_{N,k}$ theories, and the corresponding quiver diagram is as shown in figure \ref{fig:dualityb}.

One can also ask what happens for the different degeneration limits of a Riemann surface with 2 maximal punctures and $k$ minimal punctures ({\it i.e.}, the other S-dual frames of the theories in figure \ref{fig:duality}). These different limits can be described by $T_{N,k}$'s and $\cN=2$ vector and hyper multiplets.

There are a few special cases of the $T_{N,k}$'s worth mentioning, which we now go through in order of increasing $k$. First, as can be seen from the pole structure, the $T_{N,1}$ theory corresponds to $N^2$ free hypermultiplets. For $k=2$, the $T_{N,2}$ theory has its flavor symmetry enhanced to $SU(2N) \times SU(2)$.  For $k=N-1$, $T_{N,N-1}$ is identically the $T_N$, so the flavor symmetry is  enhanced to $SU(N)^3$. As can be seen from the pole structure of the third puncture, we cannot have $k>N-1$. We further note that the $T_{N,2}$, $T_{N,3}$ and $T_{N,4}$ theories feature in \cite{Chacaltana:2010ks}, where they are called $R_{0,N}$, $U_N$, and $W_N$ respectively.

We can easily obtain the graded dimensions of the Coulomb branch using equation \eqref{eq:dk}; these turn out to be $(d_2,d_3,$ $...,d_N) = \left(0,1,2,3,...,k-2,k-1,k-1,...,\right.$ $\left.k-1\right)$. We also could have calculated this by looking at the duality in figure \ref{fig:duality} and noting that $d_k$ is the same in either duality frame. Since we know $d_k$ for the linear quiver theory, we can just subtract the number of Coulomb branch operators of the superconformal tail from the whole dual theory to get $d_k$ for the $T_{N,k}$.

The central charges for these theories are
\begin{align}
&a_{T_{N,k}}=\frac{\left(6k-5\right)N^2-2k^3-\tfrac{5}{2}k^2-\tfrac{1}{2}k+5}{24}, & c_{T_{N,k}}=\frac{\left(3k-2\right)N^2-k^3-k^2+2}{12},
\label{eq:centralcharges}
\end{align}
again obtained by subtracting the central charges of the superconformal tail from those of the whole dual theory. For $k = N-1$, these recover the known expressions for $T_N$. By similar methods, we can compute the leading coefficient of the two-point function of the flavor currents, also known as the central charge $k_G$ for a flavor symmetry $G$. When we gauge a flavor symmetry, as we will do later, this quantity appears in the beta function of the associated coupling (for more info see \cite{Benini:2009mz}). For either $SU(N)$ the central charge is $k_{SU(N)}=2N$, and for $SU(k)$, the central charge is $k_{SU(k)}=2(k+1)$. 

\subsection{Higgs Branch Operators}
\label{subsec:tnkops}
Our knowledge of the Higgs branch of theories of class $\cS$ is still quite incomplete. Although some Higgs branch operators are known, and in some special cases we can make concrete statements, our knowledge of such operators is limited. In this subsection we review some relevant facts about Higgs branch operators for the $T_N$ theories, and then use similar arguments to establish the existence of analogous operators for the $T_{N,k}$ theories. We leave the much more difficult question of the structure of the Higgs branch to future work; our goal here is merely to describe some of its operators, and not the various relations between them.

The authors of \cite{Gaiotto:2009gz} give an argument for the existence of certain Higgs branch operators, which goes as follows. Consider the $SU(N)^{N-1}$ linear quiver with bifundamental hypermultiplets ({\it i.e.},~the bottom left quiver of figure \ref{fig:duality} with $n=1$). There is a gauge-invariant operator $H_{ij}=q_iA_1A_2...A_{N-2}\widetilde{q}_j$ where $q_i$ and $\widetilde{q_j}$ are the (fundamental) quarks, and the $A$'s are the bifundamentals; $i$ and $j$ are flavor indices. This operator transforms in the $\left(\textbf{N},\textbf{N}\right)$ representation of the $SU(N)^2$ flavor symmetry and has dimension $N$.  In the dual frame where the $T_N$ is coupled to a superconformal tail, this operator can be written as $H_{ij}=\cO_{ijk}Q^k$, where $Q^k$ is the quark that transforms in the fundamental representation of the $SU(N)$ gauge group and $\cO_{ijk}$ is an dimension-$(N-1)$ operator in the $\left(\textbf{N},\textbf{N},\textbf{N}\right)$ representation of the $SU(N)^3$ flavor symmetry of the $T_N$. This trifundamental is one of the Higgs branch operators in the $T_N$ theory, and a similar tri-antifundamental operator exists as well.

It is also worth considering what happens in linear quivers with different numbers of nodes. First, consider the $SU(N)^{N-2}$ linear quiver. Here, the gauge-invariant operator of interest is $H_{ij}=q_iA_1A_2...A_{N-3}\widetilde{q}_j$, which has dimension $N-1$. The dual frame corresponds to a $T_N$ where one of the $SU(N)\subset SU(N)^3$ flavor symmetries has an $SU(N-1)$ subgroup gauged. In this case, also discussed in \cite{Gaiotto:2009gz}, the operator $H_{ij}$ can be identified with $\cO_{ijN}$, which is the part of $\cO_{ijk}$ which transforms as a singlet under the $SU(N-1)$ gauge group. For the dualities denoted by $k=N-1+n$ in figure \ref{fig:duality}, 
the analogous operator can be written in the dual frame as $H_{ij}=\cO_{ijk}(A_1A_2...A_{n-1}Q)^k$, where we have hidden most gauge group indices.

As is well known, the existence of the $\cO_{ijk}$ operators in the $T_N$ theory explains the enhancement of the $SU(N)^3$ flavor symmetry to $E_6$ for the case of $N=3$. In this case, the $\cO_{ijk}$ operator is dimension two and contains a conserved current in its multiplet. Because the adjoint representation of $E_6$ decomposes under $E_6 \rightarrow SU(3)^3$ as
\begin{equation}
\textbf{78} \rightarrow \left(\textbf{8},\textbf{1},\textbf{1}\right) \oplus \left(\textbf{1},\textbf{8},\textbf{1}\right) \oplus \left(\textbf{1},\textbf{1},\textbf{8}\right) \oplus \left(\textbf{3},\textbf{3},\textbf{3}\right) \oplus \left(\overline{\textbf{3}},\overline{\textbf{3}},\overline{\textbf{3}}\right),
\end{equation}
we see that the operators $\cO^{ijk}$ (and $\overline{\cO}_{\bar i \bar j \bar k}$) combine with the currents of the $SU(3)^3$ flavor symmetry to lift the symmetry to $E_6$.

We can similarly argue for the existence of certain Higgs branch operators in the $T_{N,k}$ theories. If we look at the $SU(N)^{k-1}$ linear quiver, there is a dimension $k$ operator $H_{ij}=q_iA_1...A_{k-2}\widetilde{q}_j$ that transforms in the $\left(\textbf{N},\textbf{N}\right)$ representation of the $SU(N)^2$ flavor symmetry. As described above, for the $T_N$, there are two arguments for the existence of the operators $\cO_{ijk}$, one relying on the existence of a quark in the dual theory (when $k=N$) and one relying on a subgroup of the $SU(N)$ flavor symmetry being gauged (when $k=N-1$). However, for $k<N-1$ there is no quark in the dual theory, nor is there a gauged subgroup of the $SU(k)$. Instead we argue that the dual operator is an operator $H_{ij}=\cO_{ij}$ that transforms in the $\left(\textbf{N},\textbf{N},\textbf{1}\right)$ of the $SU(N)^2 \times SU(k)$ flavor symmetry of the $T_{N,k}$ and has dimension $k$.

One can easily see that this is the case for $k=1$, where the $T_{N,1}$ corresponds to free hypermultiplets. In this case the operator $\cO_{ij}$ is dimension one, and corresponds to the free hypermultiplets themselves. When $k=2$, these operators are dimension two. Since this case has an enhanced flavor symmetry, from $SU(N)^2 \times SU(2) \times U(1)$ to $SU(2N) \times SU(2)$, we expect that the $\cO_{ij}$ has in its multiplet the conserved currents necessary to exhibit this enhancement. The adjoint representation of $SU(2N)$ decomposes under $SU(2N) \rightarrow SU(N)^2 \times U(1)$ as
\begin{equation}
\mathbf{4N^2-1} \rightarrow \left(\mathbf{N^2-1},\mathbf{1}\right)_0 \oplus \left(\mathbf{1},\mathbf{N^2-1}\right)_0 \oplus \left(\mathbf{1},\mathbf{1}\right)_0 \oplus \left(\mathbf{N},\mathbf{\overline{N}}\right)_2 \oplus \left(\mathbf{\overline{N}},\mathbf{N}\right)_{-2},
\end{equation}
so the $\cO_{ij}$ and $\overline{\cO}^{ij}$ are exactly what we need to enhance the flavor symmetry. In the next section, we will further bolster the case for the existence of these operators by showing that the $\cO_{ij}$ appear in the superconformal index.

Along with the $\cO_{ijk}$, the $T_N$ also contains three dimension-two Higgs branch operators  $\mu_i$, $i=1,2,3$, that transform in the adjoint representation of each $SU(N)$. These operators are necessarily coupled to any relevant vector multiplets for gauged flavor symmetries via a superpotential $W = \mu \Phi$, as required by $\cN=2$ SUSY. This superpotential term is the analog of the $Q\Phi\widetilde{Q}$ term in $\cN=2$ theories with weakly coupled matter. Since the $T_{N,k}$'s also appear as part of $\cN=2$ superconformal field theories, they should also contain operators $\mu_i$, $i=1,2,3$ that transform in the adjoint representations of $SU(N)_1$, $SU(N)_2$ and $SU(k)$.

\section{The Superconformal Index}
\label{sec:SCI}

In this section we review the technology of the superconformal index \cite{Romelsberger:2005eg, Kinney:2005ej}, which we then use as a way of understanding some properties of the $T_{N,k}$ theories. Our main tool is the reduced index for type $A$ theories of class $\cS$ found in \cite{Gadde:2011ik}.

\subsection{Index Basics}
\label{subsec:indexbasics}
The $\cN=2$ superconformal index is defined as \cite{Romelsberger:2005eg,Kinney:2005ej}
\begin{equation}
\cI = \Tr \left(-1\right)^F p^{\frac{E-R}{2}+j_1} q^{\frac{E-R}{2}-j_1}u^{-(r+R)},
\end{equation}
where $F$ is the fermion number, $E$ is the conformal dimension, $R$ is the charge under the Cartan subgroup of the $SU(2)_R$ symmetry, $r$ is the charge under the $U(1)_r$ symmetry, and $(j_1,j_2)$ are the charges under the $SU(2)_1 \times SU(2)_2$ Lorentz group. $p$, $q$ and $r$ are fugacities which keep track of the quantum numbers for each state in the theory, and the trace is over states on $S^3$ in the usual radial quantization. Only states which satisfy the relationship
\begin{equation}
E - 2j_2 - 2R + r =0
\end{equation}
contribute to the index.

To help get a feel for this technology, it is useful to compute the ``single letter" contributions $f(p,q,u)$. These are the contributions to the index from all single-field operators with arbitrary numbers of derivatives. For vectors and half-hypers, the single-letter partition functions are given by (see e.g.~\cite{Gadde:2010te})
\begin{equation}
\label{singleletter}
f_{\half hyper} = \frac{(pq)^{1/4}u^{-1/2} - (pq)^{3/4}u^{1/2}}{(1-p)(1-q)}, \qquad f_{vec} = \frac{(u - u^{-1})(pq)^{1/2} - (p+q) + 2pq}{(1-p)(1-q)}.
\end{equation}
The interesting-looking 2 in the numerator of $f_{vec}$ comes from including a wrong-statistics state with the quantum numbers of a particular equation of motion. Said another way, this term subtracts contributions from states proportional to the quantity which is identified with zero by the equation of motion. 

The index in which we will be interested here, the ``reduced'' index, is obtained by setting $p=q$ and $u=1$, resulting in
\begin{equation}
\cI = \Tr \left(-1\right)^F q^{E-R}.
\end{equation}
It is easy to see that the reduced single-letter partition functions for vectors and hypers are given by
\begin{equation}
f_{\half hyper, red} = \frac{q^{1/2}}{1-q} \qquad f_{vec,red} = \frac{-2q}{1-q}.
\end{equation}
When flavor symmetries are present, extra fugacities can be introduced to keep track of the charges under the flavor symmetry. This is done in the next section.
\subsection{The 4d Superconformal Index from \texorpdfstring{$q$}{q}-deformed 2d Yang-Mills}
\label{subsec:qYM}

In \cite{Gadde:2011ik} it was conjectured that for the theories that appear in \cite{Chacaltana:2010ks},  the reduced index can be obtained using a relation to two-dimensional $q$-deformed Yang-Mills. 
Using this relationship, the index for $T_N$ is conjectured to be
\begin{equation}
\cI_{T_N} \left(\mathbf{x}_i,q\right) = \frac{\left[\Pi_{i=1}^{\infty}\left(1-q^i\right)\right]^{N-1}\left[\Pi_{i=1}^{3}\eta^{-\frac{1}{2}}\left(\mathbf{x}_i \right)\right]}{\Pi_{\ell =1}^{N-1}\left(1-q^{\ell}\right)^{N-\ell}} \left[\sum_{\cR} \frac{1}{\dim _q \cR} \chi_{\cR} \left(\mathbf{x}_1\right) \chi_{\cR} \left(\mathbf{x}_2\right) \chi_{\cR} \left(\mathbf{x}_3\right)\right].
\label{eq:scTN1}
\end{equation}
The fugacities in $\cI$ are $q$ (which keeps track of the $E$ and $R$ charges), as well as the vectors $\mathbf{x}_i, i=1,2,3$, which are associated to the three punctures (and keep track of the charges under the flavor symmetry); we will go into greater detail about these below.
The sum in \eqref{eq:scTN1} is over irreducible representations of $SU(N)$, and the $q$-deformed dimension of a representation is given by
\begin{equation}
\dim _q \cR = \prod _{i<j} \frac{\left[\lambda_i - \lambda_j + j -i\right]_q}{\left[j-i\right]_q},
\end{equation}
where $\lambda_1 \geq \lambda_2 \geq ... \geq \lambda_{N-1} \geq \lambda_N=0 $ are the row lengths of the Young tableau corresponding to the representation, $\cR$, and a $q$-deformed number $[x]_q$ is defined as
\begin{equation}
\left[ x \right]_q = \frac{q^{-\frac{x}{2}}-q^{\frac{x}{2}}}{q^{-\frac{1}{2}}-q^{\frac{1}{2}}}.
\end{equation}
The characters in equation \eqref{eq:scTN1} are given by the Schur polynomials:
\begin{equation}
\chi_{\cR} \left( \mathbf{x} \right) = \frac{\det\left(x_i^{\lambda_j +N -j}\right)}{\det\left(x_i^{N-j}\right)},
\label{eq:char}
\end{equation}
where {\it e.g.} $x_i^{N-j}$ is to be thought of as the entry in the $i^{\rm th}$ row and $j^{\rm th}$ column of a matrix.
Finally, the quantity $\eta(\mathbf{x})$ is given by
\begin{equation}
\eta \left(\mathbf{x}\right) = \exp \left\{ -2\sum_{n=1}^{\infty} \frac{1}{n} \frac{q^n}{1-q^n} \chi_{\textbf{Adj}} \left(\mathbf{x}^n\right) \right\}.
\end{equation}

\begin{figure}[ht]\begin{center}
\includegraphics[scale=0.3]{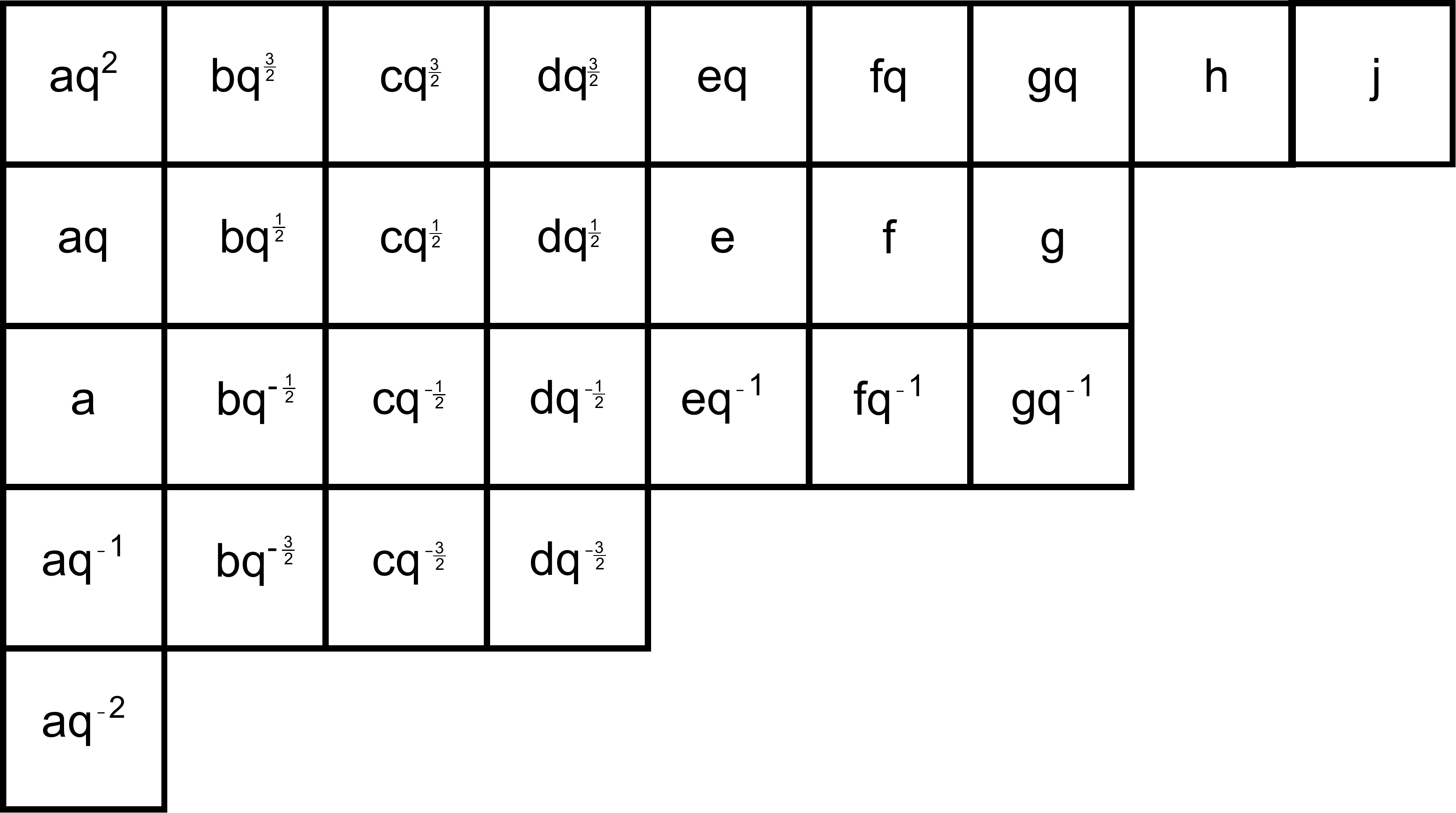}
\caption{An example of the association of flavor fugacities to a Young tableau. 
The flavor symmetry associated to this puncture is $S\left(U(3)^2 \times U(2) \times U(1) \right)$ $=SU(3)^2 \times SU(2) \times U(1)^3$ and the $S(...)$ constraint imposes $a^5(bcd)^4(efg)^3hj=1$.}
\label{fig:fug}
\end{center}\end{figure}

To get the index for a more general fixture, rather than one with only maximal punctures, we must first associate flavor fugacities to each puncture using the prescription outlined in \cite{Gadde:2011ik}. The prescription is as follows: Take the Young tableau associated to the puncture, and associate a fugacity to each column of the tableau. For each box in the tableau, associate the fugacity for that column times some power of $q$. The powers of $q$ should decrease by one down each column and be symmetric about 0; a column with $n$ boxes will begin with the power $q^{(n-1)/2}$. Finally, impose the condition that the product of the quantities associated to each box in the tableau equals 1. This procedure is exemplified in figure \ref{fig:fug}.
The conjecture then is that the index for a fixture is given by
\begin{equation}
\cI (\mathbf{x}_i) = \cN(q) \left[\prod_{i=1}^3 \cA(\mathbf{x}_i)\right] \sum_{\cR} \frac{1}{\dim _q \cR} \chi _{\cR}(\mathbf{x}_1) \chi _{\cR}(\mathbf{x}_2) \chi _{\cR}(\mathbf{x}_3),
\end{equation}
where $\cN$ and $\cA$ are normalization factors associated to the fixture and punctures, respectively. For the case of a maximal puncture, $\cA(\mathbf{x}) = \eta (\mathbf{x})$. 

As a warm-up to the $T_{N,k}$, we can expand the expression \eqref{eq:scTN1} to get some useful information. It is useful to note before we begin that each factor goes to 1 as $q$ goes to zero, so it is easy to read off the low powers of $q$.
First, since
\begin{equation}
\eta^{-\frac{1}{2}}(\mathbf{x}) = 1 + q\, \chi_{\textbf{Adj}} (\mathbf{x}) + \cO \left(q^2 \right),
\end{equation}
we see a contribution to $\cI_{T_N}$ of the form $q \sum_{i=1}^3 \chi_{\textbf{Adj}} (\mathbf{x_i})$. These terms represent the dimension-2 Higgs branch operators $\mu_i$ in the adjoint of each $SU(N)$ flavor symmetry of the $T_N$.

Now consider the terms with $\cR=\fund,\afund$. Because $\dim _q \afund = \dim _q \fund = \left [ N \right ]_q =  q^{\frac{-(N-1)}{2}} (1+\cO(q))$, there is a term of the form
\begin{equation}
q^{\frac{N-1}{2}} \left[ \chi_{\fund} \left(\mathbf{x}_1\right) \chi_{\fund} \left(\mathbf{x}_2\right) \chi_{\fund} \left(\mathbf{x}_3\right) +  \chi_{\afund} \left(\mathbf{x}_1\right) \chi_{\afund} \left(\mathbf{x}_2\right) \chi_{\afund} \left(\mathbf{x}_3\right) \right].
\end{equation}
This  is the contribution from the dimension-$(N-1)$ operators $\cO_{ijk}$ and $\overline{\cO}^{ijk}$ , which are in the trifundamental and tri-antifundamental representations. 

Finally,  we look at the term that comes from $\cR = \Lambda^l$, the $l$-index fully antisymmetric representation. Since
\begin{equation}
\frac{1}{\dim _q \Lambda^l} = q^{\frac{l}{2}(N-l)} \left(1 + \cO(q)\right),
\label{eq:dimlambdal}
\end{equation}
there will be a term of the form $q^{\frac{l}{2}(N-l)}\left(\chi_{\Lambda^l}(\mathbf{x_1}) \chi_{\Lambda^l}(\mathbf{x_2}) \chi_{\Lambda^l}(\mathbf{x_3})\right)$, indicating the presence of dimension-$l(N-l)$ operators in the $\left(\Lambda^l,\Lambda^l,\Lambda^l\right)$ representation. It is interesting to note that at present it is not known how to get these operators with $l \neq 1, N-1$ from duality arguments.

\subsection{The Superconformal Index for the \texorpdfstring{$T_{N,k}$}{TNk}}
\label{subsec:sctnk}
We now look at calculating the reduced superconformal index for the $T_{N,k}$ theory. The $T_{N,k}$ theory has two maximal punctures and one puncture with pole structure $\{1,2,...,k-1,k,k,...k\}$, so the superconformal index is
\begin{align}
\cI_{T_{N,k}} (\mathbf{x}_i) = \cN_{T_{N,k}}(q) \left[\prod_{i=1}^2 \eta^{-\frac{1}{2}} (\mathbf{x}_i) \right] \cA(\mathbf{x}_3) \nonumber \left[ \sum_{\cR} \frac{\chi_{\cR} (\mathbf{x}_1) \chi_{\cR} (\mathbf{x}_2) \chi_{\cR} \left( \mathbf{x_3} \right)}{\dim _q \cR} \right].
\label{eq:sctnk}
\end{align}
The Young tableau for the non-maximal puncture has one column of height $N-k$ and $k$ columns of height one, giving the flavor fugacities
\begin{equation}
\mathbf{x_3} = \left( aq^{\frac{N-k-1}{2}},...,aq^{-\frac{N-k-1}{2}}, b_1a^{\frac{k-N}{k}},b_2a^{\frac{k-N}{k}},...,b_{k-1}a^{\frac{k-N}{k}},\left[\Pi_{i=1}^{k-1}b_i\right]^{-1}a^{\frac{k-N}{k}} \right).
\end{equation}

We first look at the terms in the last factor with $\cR=\fund, \afund$.
Eq.~\eqref{eq:char} gives us $\chi_{\Box}(\mathbf{x})=\sum_i x_i$, so the characters of the third puncture are given by 
\begin{align}
&\chi_{\fund} \left( \mathbf{x}_3 \right) = aq^{-\frac{N-k-1}{2}}+...+aq^{\frac{N-k-1}{2}}+a^{\frac{k-N}{k}}\chi_{\fund}\left(\mathbf{b}\right),
\end{align}
with $\chi_{\afund}$ given by taking $a \rightarrow 1/a$ and $\fund \rightarrow \afund$.
$\chi_{\fund}\left(\mathbf{b}\right)$ and $\chi_{\afund}\left(\mathbf{b}\right)$ are the characters of the fundamental and anti-fundamental representations of $SU(k)$ in terms of the flavor fugacities $b_1,...,b_{k-1}$. Multiplying by the factors from the maximal punctures, $q^{\frac{1}{2}(N-1)} \chi_{\fund} (\mathbf{x}_1) \chi_{\fund} (\mathbf{x}_2)$ and $q^{\frac{1}{2}(N-1)} \chi_{\afund} (\mathbf{x}_1) \chi_{\afund} (\mathbf{x}_2)$ respectively, we see the presence of operators given in table \ref{tab:tnkfundops}. In addition to the (perhaps expected) presence of trifundamental and tri-antifundamental operators, it is interesting to note the presence of bifundamentals of various dimensions increasing in increments of two including the dimension $k$ bifundamental operators found in the last section.
\begin{table}[ht]\begin{center}
\begin{tabular}{c c c}
No of Operators & Representation & Dimension \\
\hline
$N-k$ & $\left(\fund,\fund, \mathbf{1}\right)_1$ & $k,k+2,k+4,...,2N-k-2$ \\
$N-k$ & $(\afund,\afund, \mathbf{1})_{-1}$ & $k,k+2,k+4,...,2N-k-2$ \\
$1$&$\left(\fund,\fund, \fund\right)_{\frac{k-N}{k}}$ & $N-1$ \\
$1$&$(\afund,\afund, \afund)_{\frac{N-k}{k}}$ & $N-1$ \\
\end{tabular}
\caption{The flavor symmetry representations (under $SU(N)^2 \times SU(k) \times U(1)$) and dimensions of operators for the $T_{N,k}$.}
\label{tab:tnkfundops}
\end{center}\end{table}

Note that the number of operators coming from the $\cR=\fund$ or $\cR=\afund$ terms is $N^2(N-k+k)=N^3$, which is the same counting as in the analogous terms for the $T_N$. This has to be the case since $\chi \left( \mathbf{x}_3 \right)$ has the same number of terms regardless of the puncture. Moreover, when $k=N-1$, the dimension-$(N-1)$ bifundamental and trifundamental operators combine to give us the trifundamental operator of the $T_N$, as expected.

It is also interesting to consider the operators transforming in the various $\ell$-index antisymmetric representations $\Lambda^\ell$. Using equation \eqref{eq:char} one can see that the characters of these representations are given by
\begin{equation}
\chi_{\Lambda^l}(\mathbf{x}) =  \sum_{\substack{\text{$i_1=1,$}\\ \text{$i_2>i_1,...,i_{l}>i_{l-1}$}}}^{N}x_{i_1}x_{i_2}...x_{i_{l}},
\end{equation}
and so for the third puncture we have
\begin{equation}
\chi_{\Lambda^l}(\mathbf{x_3}) = \sum_{l'=0}^{\min(l,k)}a^{l-l'\left(\frac{N}{k}\right)}\left[ \sum_{\substack{\text{$i_1=-\frac{N-k-1}{2},$}\\ \text{$i_2>i_1,...,i_{l-l'}>i_{l-l'-1}$}}}^{\frac{N-k-1}{2}}q^{i_1}q^{i_2}...q^{i_{l-l'}} \right] \chi_{\Lambda^{l'}}\left(\mathbf{b}\right).
\end{equation}
We find that the operators coming from the $\cR = \Lambda^l$ term are those given in table \ref{tab:tnklambdalops}. Again we see that the number of operators coming from the $\cR = \Lambda^l$ term is equal to that of the $T_N$, which we can see via Vandermonde's identity:
\begin{equation}
\sum_{l'=0}^{\min(l,k)}\begin{pmatrix} N-k \\ l-l' \end{pmatrix}\begin{pmatrix} k \\ l' \end{pmatrix} = \begin{pmatrix} N \\ l \end{pmatrix}.
\end{equation}
\begin{table}[ht]\begin{center}
\begin{tabular}{c c c}
No of Operators & Representation & Dimension \\
\hline
$\begin{pmatrix} N-k \\ l \end{pmatrix}$ & $\left(\Lambda^l,\Lambda^l,\mathbf{1}\right)_{l}$ & $[lk,2lN-k l-2 l^2]$ \\
$\begin{pmatrix} N-k \\ l-1 \end{pmatrix}$ & $\left(\Lambda^l,\Lambda^l,\fund\right)_{l-\left(\frac{N}{k}\right)}$ & $\substack{\text{$[N+kl-2l-k-1,$}\\ \text{$2lN-N-2l^2-kl+2l+k-1]$}}$ \\
\vdots & \vdots & \vdots \\
$\begin{pmatrix} N-k \\ l-l' \end{pmatrix}$ & $\left(\Lambda^l,\Lambda^l,\Lambda^{l'}\right)_{l-l'\left(\frac{N}{k}\right)}$ & $\substack{\text{$[lk+l'N-l'k-2ll'+l'^2,$}\\ \text{$(k+2l)(l'-l)+N(2l-l')-l'^2]$}}$ \\
\end{tabular}

\caption{This table gives the flavor symmetry representations (under $SU(N)^2 \times SU(k) \times U(1)$) and range of dimensions of operators of the $T_{N,k}$. $l'$ will stop at $l$ or $k$, whichever is less.}
\label{tab:tnklambdalops}
\end{center}\end{table}


\section{New SCFTs and Flows}
\label{sec:flows}
In this section we use the $T_{N,k}$ theories to construct new $\cN=1$ SCFTs, and describe flows between these theories. The analysis in this section extends the work done in \cite{Bah:2011je} and answers an open question about flows that appeared to violate the $a$-theorem. We note here that the evidence presented in this and the following section is necessary but not sufficient for the theories in question to be SCFTs. Although some of the theories we will build have obvious problems such as unitarity violations, it is possible that even the ones that do not appear to be problematic do not actually dynamically reach a conformal fixed point. To determine without question whether or not the theories we consider are conformal would require stronger evidence, such as an AdS dual. Nevertheless, we believe the evidence presented here is suggestive that many of these theories are SCFTs.

\subsection{\texorpdfstring{$S_{\ell}$}{Sl} Theories With \texorpdfstring{$T_N$}{TN}'s}
\label{subsec:sltn}
\begin{figure}[ht]\begin{center}
\includegraphics[scale=.15]{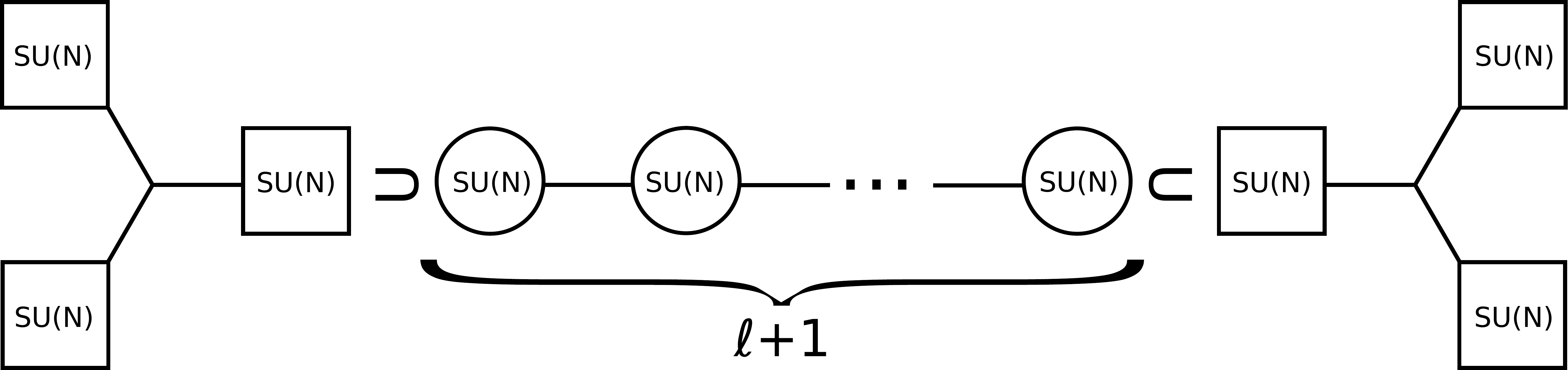}
\caption{The $S_{\ell}$ quiver.}
\label{fig:sltn}
\end{center}\end{figure}

An $S_{\ell}$ theory, first analyzed in \cite{Bah:2011je}, is an $\cN=1$ $SU(N)^{\ell+1}$ gauge theory with $\ell$ bifundamental hypermultiplets, two $T_N$'s, and an $SU(N)^4 \times U(1) \times U(1)_R$ global anomaly-free symmetry. The theory is represented by the generalized quiver shown in figure \ref{fig:sltn}.
Since we are now dealing with $\cN=1$ theories, in this section circles will correspond to $\cN=1$ vector multiplets. Lines will still correspond to bifundamental $\cN=2$ hypermultiplets, or in $\cN=1$ language, two chiral multiplets in the $(\fund, \afund)$ and $(\afund,\fund)$ representations.

A useful global symmetry is 
\begin{equation}
R_0 = R_{\cN=1}+\frac{1}{6}J = \frac{1}{2}R_{\cN=2}+I_3,
\label{eq:r0}
\end{equation}
which is the $R$-symmetry preserved when flowing to the $S_\ell$ theory by giving masses to adjoint chiral superfields in vector multiplets in the analogous $\cN=2$ theory. 
The $U(1)$ global symmetries $R_{\cN=2}, R_{\cN=1}, J,$ and $I_3$ are detailed in \cite{Bah:2011je}. Here we only note that $R_{\cN=2}$ and $I_3$ are the charges under the $U(1)_R \times U(1)_r$ which descends from the $U(1)_R \times SU(2)_r$ $\cN=2$ $R$-symmetry; $R_{\cN=1}$ and $J$ are just particular linear combinations. Additionally, each of the bifundamentals comes with a $U(1)$ which we call $F_i$, normalized as $F_i(Q_j)=F_i(\widetilde{Q}_j)=\delta_{ij}$ where $Q_j, \widetilde Q_j$ are the $j^{\rm th}$ bifundamentals. These $F_i$ are individually anomalous but can be combined into the anomaly-free global symmetry
\begin{equation}
\cF = J_1 + \sum_{i=1}^{\ell} (-1)^{i-1}F_i + (-1)^{\ell-1}J_2,
\label{eq:addu1}
\end{equation}
where $J_{1,2}$ are global symmetries under which only the $T_N$ theories are charged. When $\ell$ is even, $\Tr \cF=0$, so $\cF$ will not mix with the $R$-symmetry \cite{Intriligator:2003jj}, which is therefore $R_0$.

When $\ell$ is odd, $\Tr \cF \neq 0$, so we must use $a$-maximization \cite{Intriligator:2003jj} to determine the IR $R$-symmetry. In other words, we must find the value of $\alpha$ that maximizes
\begin{equation}
a_{trial}( \alpha )=3\Tr R_{trial}^3 - \Tr R_{trial},
\end{equation}
where $R_{trial}(\alpha)=R_0+\alpha \cF$. This was done in \cite{Bah:2011je}, with the result that
\begin{equation}
\widehat{\alpha}=\frac{A- \sqrt{B}}{C},
\end{equation}
where
\begin{align}
&A=4N^3+3\ell N^2-4N, \nonumber \\
&B=64N^6 +8\left(3\ell-25\right)N^5 +3\left(3\ell^2+41\right)N^4 -24\left(\ell-9\right)N^3 -208N^2 -64N +64, \nonumber \\
&C=6\left(4N^3-11N^2+8\right). \nonumber
\end{align}

\subsection{\texorpdfstring{$S_{\ell}$}{Sl} Theories With \texorpdfstring{$T_{N,k}$}{TNk}'s}
\label{subsec:sltnk}
We now look at the $S_{\ell}$ theory as in the last section but now with $T_{N,k}$'s and an $SU(N) \subset SU(N)^2 \times SU(k) \times U(1)$ gauged at each end of the quiver. We again find that there is an anomaly-free $R$-symmetry as in equation \eqref{eq:r0} and an anomaly-free $U(1)$ flavor symmetry as in equation \eqref{eq:addu1}. As before, the case with even $\ell$ is trivial, and the $R$-symmetry is $R_{0}$. However, for $\ell$ odd, we must use $a$-maximization. 

\begin{figure}[ht]\begin{center}
\includegraphics[scale=.4]{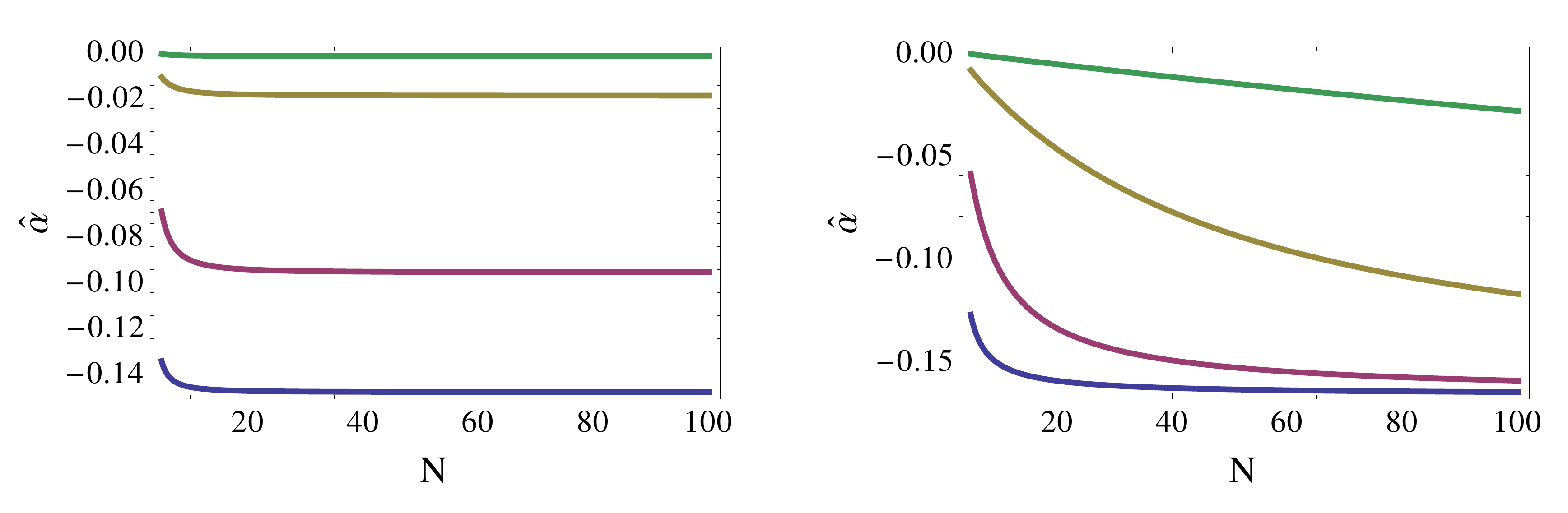}
\caption{The value of $\alpha$ that maximizes $a_{trial}$ for the $S_{\ell}$ theory with two $T_{N,k}$'s. $\alpha$ is plotted against $N$ for $k=5$ (left), and $k=N-2$ (right) each with $\ell = 1$ (blue), $11$ (purple), $101$ (yellow), $1001$ (green). }\label{fig:sltnkalpha}
\end{center}\end{figure}

If we perform $a$-maximization then we find that the value of $\alpha$ that maximizes $a_{trial}$ is
\begin{equation}
\widehat{\alpha}=\frac{A+ \sqrt{B}}{C},
\end{equation}
where
\begin{align}
A=&\ -(3\ell+6k)N^2 +2k^3-2k, \nonumber \\
B=&\ N^4 \left(144 k^2+36 k \ell-204 k+9 \ell^2+91 \right) \nonumber \\
&+N^2 \left(-96 k^4-12 k^3 \ell-28 k^3+80 k^2+12 k \ell+204 k-160\right) \nonumber \\
&+16 k^6+32 k^5+16 k^4-64 k^3-64 k^2+64, \nonumber \\
C=&\ 6\left((7-6 k) N^2+2 k^3+4 k^2+2 k-8\right). \nonumber
\end{align}
$\widehat{\alpha}$, which is plotted in figure \ref{fig:sltnkalpha}, seems to be negative for all values of $\ell$, $N$ and $k$ and approaches $\frac{-6 k-3 \ell +\sqrt{144 k^2-204 k+36 k\ell +9 \ell ^2+91 }}{6 (7-6 k)}$ at large $N$.

\begin{wraptable}{r}{0.35\textwidth}\begin{center}
\begin{tabular}{c c}
Operator & $R$-charge \\
\hline
$Q_i$ & $\frac{1}{2}+(-1)^{i-1}\alpha$ \\
$\widetilde{Q}^i$ & $\frac{1}{2}+(-1)^{i-1}\alpha$ \\
$\mu$ & $1-2\alpha$ \\
$\cO_{H}$ & $\left(\frac{1}{2}-\alpha\right)\Delta_{UV}$ \\
$u_n$ & $\left(1+2\alpha\right)n$ \\
\end{tabular}
\caption{Operator dimensions for the $S_{\ell}$ theory with $T_{N,k}$'s. $\cO_H$ is any of the Higgs branch operators in table \ref{tab:tnklambdalops} and $\Delta_{UV}$ is the operator's dimension given in the same table. $u_n$ are the Coulomb branch operators.}
\label{tab:sltnkopdims}
\end{center}\end{wraptable}

In order to verify that there are no gauge-invariant operators in this theory that violate the unitarity bound $R \geq \frac{2}{3}$ we note that $\alpha$ never goes below $-\frac{1}{6}$ for any $k$ and $\ell$. One can then easily verify using the results of table \ref{tab:sltnkopdims} that indeed no gauge-invariant operators violate unitarity.

We can also ask what happens when we construct the $S_{\ell}$ theory with two different $T_{N,k}$'s at either end of the quiver, {\it i.e.}, $T_{N,k_1}$ and $T_{N,k_2}$. The behavior is qualitatively similar to when $k_1=k_2$, and we have included the result in the appendix. For now, we merely note that no gauge-invariant operators violate the unitarity bound, so these theories do not appear to be problematic.

\subsection{Other \texorpdfstring{$\cN=1$}{N=1} Theories With \texorpdfstring{$T_{N,k}$}{TNk}'s}
\label{subsec:othn1}
\begin{figure}[ht]\begin{center}
\includegraphics[scale=0.2]{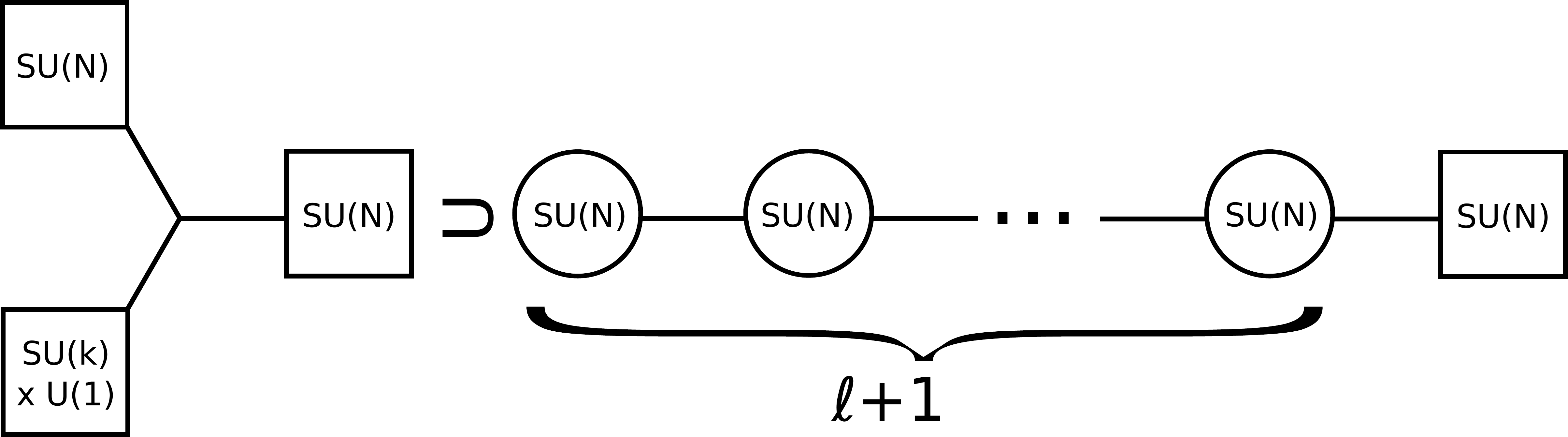}
\caption{The generalized quiver diagram for an $S^{\Box}_{\ell}$ theory.}
\label{fig:slboxtn}
\end{center}\end{figure}

\begin{figure}[ht]\begin{center}
\includegraphics[scale=0.2]{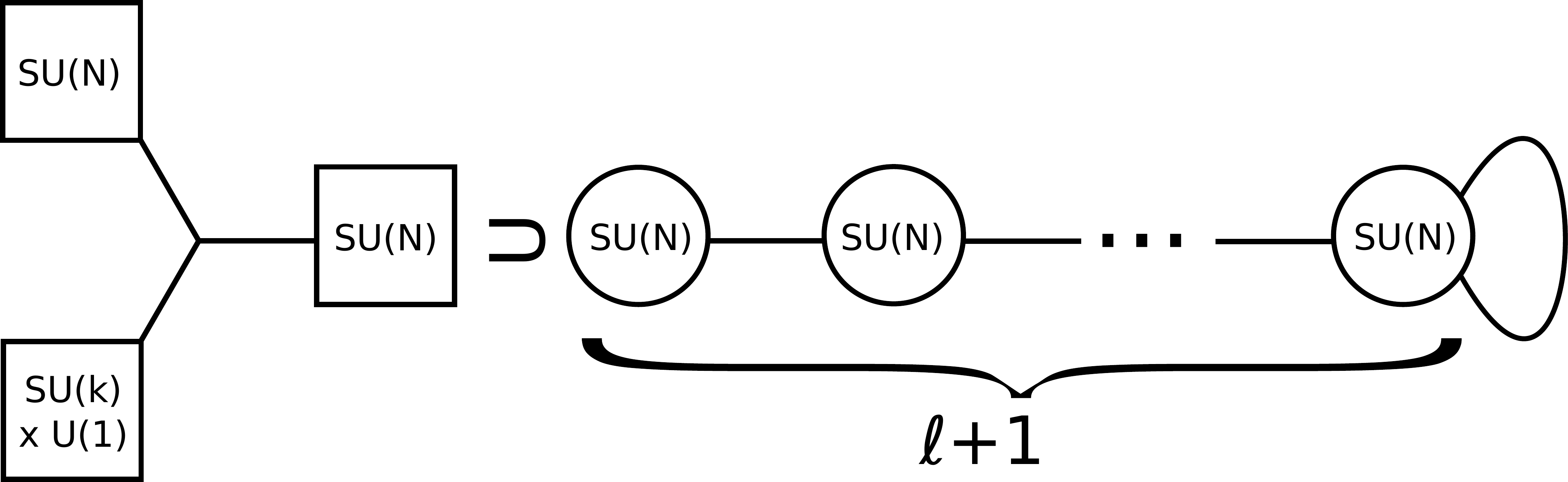}
\caption{The generalized quiver diagram for an $S^{\circ}_{\ell}$ theory. The loop denotes a chiral superfield in the adjoint representation.}
\label{fig:slcirctn}
\end{center}\end{figure}

In \cite{Bah:2011je}, the authors additionally studied two other theories formed from $T_N$'s and Lagrangian matter, namely the $S^{\Box}_{\ell}$ and the $S^{\circ}_{\ell}$. We now wish to construct analogs of the theories using $T_{N,k}$'s instead of $T_N$'s. The generalized quiver diagrams for these theories are given in figures \ref{fig:slboxtn} and \ref{fig:slcirctn}.


The extension of the $S^{\Box}_{\ell}$ theory is straightforward, since it is a special case of the theories at the end of the previous section. Because the $T_{N,1}$ is a set of $N^2$ free hypermultiplets, the theories at the end of the previous section with $k_2=1$ are the $S^{\Box}_{\ell}$ theory.

For the $S^{\circ}_{\ell}$ theories, we use the same $R_0$ symmetry as in the previous section. In order for $R_0$ to be anomaly-free we require that $R_0(\Phi)=\frac{1}{2}$, where $\Phi$ is the adjoint chiral superfield. The extra adjoint chiral superfield $\Phi$ comes with a $U(1)$ flavor symmetry $F_a$ which we normalize so that $F_a(\Phi)=1$. The only anomaly-free $U(1)$ symmetry is

\begin{equation}
\cF = J_1 + \sum_{i=1}^{\ell}(-1)^{i-1}F_i + (-1)^{\ell}F_a
\end{equation}

Again we can get the IR $R$-symmetry by maximizing $a_{trial} (\alpha) = 3\Tr R_{trial}^3 - \Tr R_{trial}$ with respect to $\alpha$, where $R_{trial}=R_0 + \alpha \cF$. The answer is unwieldy, so we merely note that $\alpha$ does not seem to drop below $-\frac{1}{6}$ for any $\ell,k$ and consequently there are no unitarity bound violations for the same reasons as for the $S_{\ell}$ theories. Thus, the theories in this subsection are likely to be good SCFTs.

It is interesting to note that when we add the superpotential term $Q_{\ell}\Phi\widetilde{Q}_{\ell}$ to the theory, some of the operators violate unitarity. In the theory without this superpotential term the $R$-charge of the $Q_{\ell}\Phi\widetilde{Q}_{\ell}$ operator is $R(Q_{\ell}\Phi\widetilde{Q}_{\ell})=\frac{3}{2}-\left(-1\right)^{\ell}\widehat{\alpha}$, where $\widehat{\alpha}$ is the value of $\alpha$ that maximizes $a_{trial}$. There are also operators in the theory $\Tr ( \Phi^n )$ which have $R$-charge $R(\Phi^n) = n\left(\frac{1}{2}+(-1)^{\ell}\widehat{\alpha}\right)$. In the theory with the superpotential term turned on $a$-maximization is not needed because the $R$ charge of the $Q_{\ell}\Phi\widetilde{Q}_{\ell}$ term is fixed to equal 2. This effectively sets the value of $\widehat{\alpha}$ so that $(-1)^{\ell}\widehat{\alpha}=-\frac{1}{2}$. This means that the $R$-charge of the  $\Tr ( \Phi^n )$ operators will be zero. Thus, these theories with the superpotential term turned on appear to be problematic, and are likely not SCFTs.

\subsection{Flows From Higgsing}
\label{subsec:higgs}
\begin{figure}[ht]\begin{center}
\includegraphics[scale=0.15]{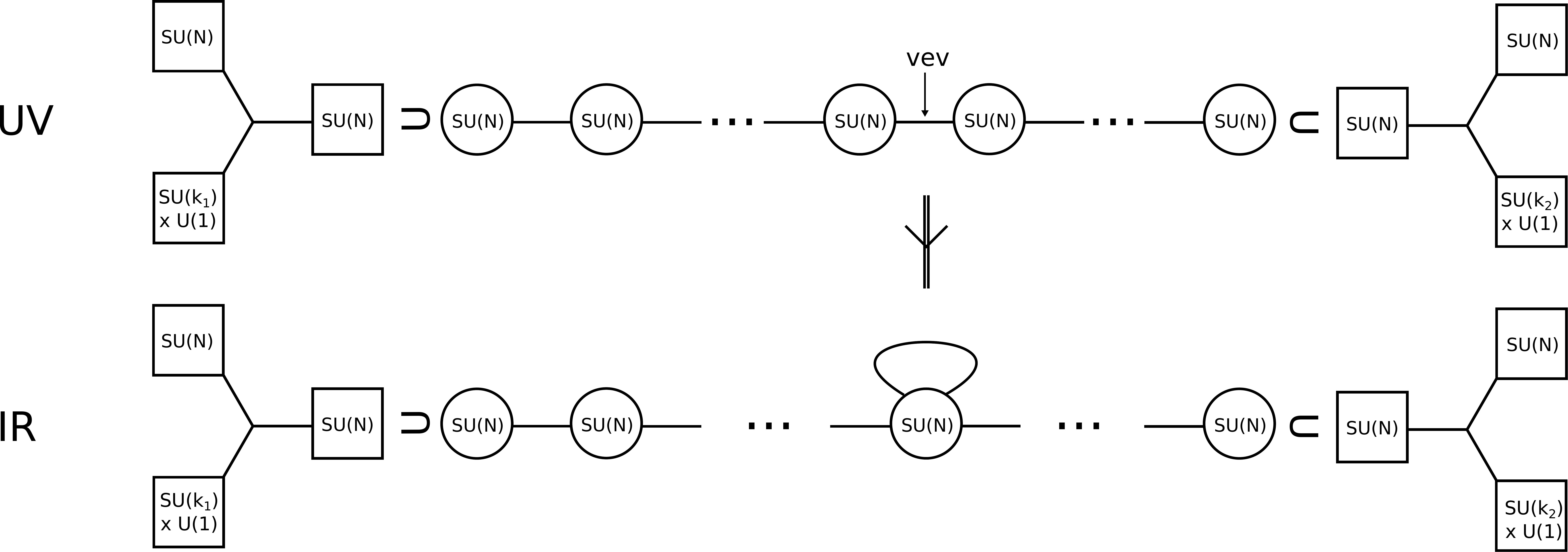}
\caption{If we give a vev to the $k$-th hypermultiplet in the $S_{\ell}$ theory (above) then this induces a flow to the $S_{\ell-1}$ theory with an chiral superfield in the adjoint representation of the $(k-1)$-th gauge group.}
\label{fig:flow}
\end{center}\end{figure}

We now look at what happens when we take an $S_{\ell}$ theory with $T_{N,k}$'s and give a vev to the $k$-th hypermultiplet. In \cite{Bah:2011je} it was argued that the theory that emerges in the IR is the $S_{\ell -1}$ theory with a chiral superfield $\Phi$ in the adjoint representation of the $(k-1)$-th gauge group \footnote{The authors of \cite{Bah:2011je} only considered the $S_{\ell}$ theory with $T_N$'s but the argument still holds.}. This is represented by the quivers in figure \ref{fig:flow}.

We note that in the UV theory there are the marginal operators $Q_{k-1}\widetilde{Q}_{k-1}Q_{k}\widetilde{Q}_{k}$ and $Q_{k}\widetilde{Q}_{k}Q_{k+1}\widetilde{Q}_{k+1}$, so we must in general consider these terms to be turned on. After Higgsing, these terms become $Q_{k-1}\Phi\widetilde{Q}_{k-1}$ and $Q_{k+1}\Phi\widetilde{Q}_{k+1}$ respectively, where $\Phi$ is the adjoint chiral superfield. These superpotential terms were not taken into consideration in \cite{Bah:2011je}, and since they are allowed by all symmetries, should in general be included.\footnote{A similar point was discussed in \cite{Bah:2013aha}.}

Including such terms results in the one-parameter family of R-symmetries
\begin{equation}
R_{trial}=R_0+\alpha \cF,
\end{equation}
where the additional anomaly-free $U(1)$ symmetry is
\begin{align}
\cF =& J_1+F_1-F_2+ ... +(-1)^{k}F_{k-1} + 2(-1)^{k+1}F_a + (-1)^{k+2}F_{k+1}+... \nonumber \\
&+(-1)^{\ell}F_{\ell-1}+(-1)^{\ell}J_2.
\label{eq:fhiggs}
\end{align}
In this formula $F_a$ is the additional $U(1)$ symmetry that comes with the adjoint chiral superfield,  which we normalize as $F_a(\Phi)=1$. We can then use $a$-maximization to find the value of $\alpha$ that maximizes $a$; this result is again in the appendix. 

\begin{figure}[ht]\begin{center}
\includegraphics[scale=0.4]{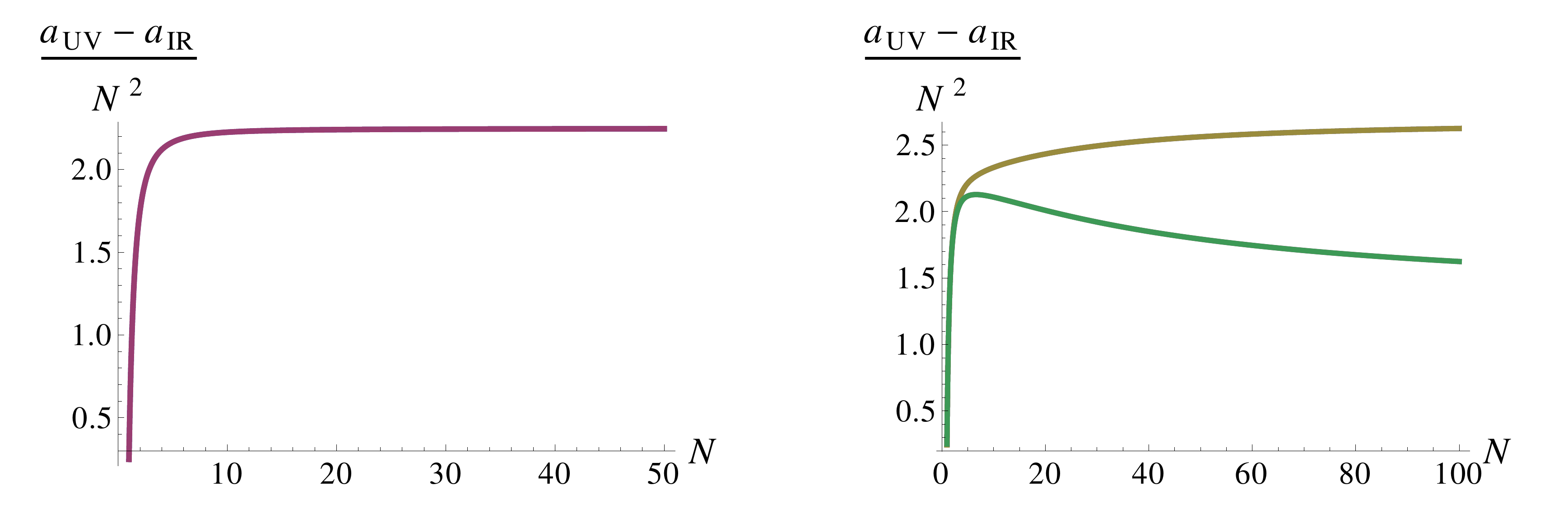}
\caption{Left: $a_{UV}-a_{IR}$, which is independent of $k$, plotted for even $\ell$. Right: $a_{UV}-a_{IR}$ plotted for $\ell$ odd and $k$ even (bottom) and odd (top).} 
\label{fig:ahiggs}
\end{center}\end{figure}

If we then calculate $a_{UV}-a_{IR}$ then we see that there are no $a$-theorem violations for this flow. The value of $a_{UV}-a_{IR}$ for even and odd $\ell$ is plotted against $N$ in figure \ref{fig:ahiggs}.
We can repeat this analysis for the $S_{\ell}$ with two general $T_{N,k_1}$, $T_{N,k_2}$ and we find that there are no $a$-theorem violations for any of these flows. Although not in and of itself conclusive, the fact that none of these flows violates the $a$-theorem lends credence to the existence of the IR theories as interacting conformal points. This is perhaps not surprising, since many examples of such quivers which mix $\cN=1$ and $\cN=2$ vector multiplets are now known to be SCFTs, though these were not known at the time of the original work \cite{Bah:2011je}.

\subsection{Linear Deformations of the \texorpdfstring{$T_{N,k}$}{TNk}}
\label{subsec:lindef}

In this section we look at what happens when we deform a $T_{N,k}$ theory with an operator of the form $\Tr (\phi \mu)$, where $\phi$ is a constant adjoint-valued matrix. This in general breaks the flavor symmetry of the $T_{N,k}$ and drives a flow to a new theory in the IR. We can use the methods of \cite{Heckman:2010qv} to determine the $R$-symmetry in the IR.

For simplicity, as was done in \cite{Heckman:2010qv}, we assume that the matrix $\phi$ takes block diagonal form, $\phi = \oplus_{a}\phi^{(a)}$, where each $\phi^{(a)}$ is an $n_a \times n_a$ upper-diagonal matrix; this breaks the theory to $\cN=1$. Then there is an $SU(2)$ subalgebra of the original flavor symmetry associated to each $\phi^{(a)}$, where $\phi^{(a)}$ is in the spin-$\frac{1}{2}(n_a-1)$ representation. As discussed in \cite{Heckman:2010qv}, the entries of $\phi^{(a)}$ along the first superdiagonal are the most relevant and drive the flow, so we further assume that each $\phi^{(a)}$ is a nilpotent Jordan block.

The IR $R$-symmetry then is given by
\begin{equation}
R_{IR} = \frac{t}{2}R_{\cN=2} + (2-t)I_3 - tT_3,
\end{equation}
where $T_3=\sum_aT_3^{(a)}$ and $T_3^{(a)}$ is the generator of the Cartan subalgebra of the $SU(2)$ flavor symmetry associated to $\phi^{(a)}$; $t$ is determined by $a$-maximization to be
\begin{equation}
t=\frac{4}{3}\times \frac{8a_{T_{N,k}}-4c_{T_{N,k}}-\sqrt{4c_{T_{N,k}}^2+(4a_{T_{N,k}}-c_{T_{N,k}})k_Gr}}{16a_{T_{N,k}}-12c_{T_{N,k}}-k_Gr},
\label{eq:t}
\end{equation}
where $k_G$ is the central charge of the flavor symmetry under which the $\mu$ we are deforming with transforms; $r=2 \Tr (T_3T_3)$ measures the sizes of the $\phi^{(a)}$ blocks. We will look only at the case where $\phi$ has just a single $n \times n$ upper-diagonal block so that $r=\frac{n^3-n}{6}$.

We can calculate the central charge $a$ for this theory in the IR and compare it to $a_{UV}$ given in eq.\eqref{eq:centralcharges}. When we do this, we see that $a_{UV}(N,k)-a_{IR}(N,k,n)$ never drops below zero, as dictated by the $a$-theorem, and is monotonically increasing in $N$, $k$ and $n$. This makes intuitive sense, because increasing $N$ or $k$ corresponds to adding degrees of freedom to the UV theory, and increasing $n$ corresponds to integrating out a larger proportion of the IR degrees of freedom.

We can also ask what the operator dimensions are in these theories. First we look at the Coulomb branch operators, which have dimension equal to $\frac{3}{2}t\Delta_{UV}$. Since the lowest lying operators have dimension $\Delta_{UV}=3$, to check if unitarity bounds are violated, it suffices to check if $t$ drops below $\frac{2}{9}$. It is easy to show that this happens for many values of $N$ and $k$ provided $n$ is large enough. Thus it seems that these are likely not good SCFTs in general, although it is possible that there is some interesting reason for critical values of $n$. In the absence of an understanding of why this transition should happen, it seems most reasonable to conclude that none of these theories are conformal.

We now look at the dimension of the $\mu$ operator. Because we deformed by a $\Tr (\phi \mu)$ operator and identified a $U(1)$ symmetry to $\phi$ (or equivalently $\mu$) we see that $\mu$ splits up into many operators with different $T_3$ charge. The dimensions of these operators are $ \frac{3}{2}\left(2-t(1+T_3)\right)$. It is easy to verify that many of these operators violate the unitarity bound, $\Delta \geq 1$, for many values of $N$, $k$ and $n$.

For the Higgs branch operators that we found using the superconformal index (i.e. those given in table \ref{tab:tnklambdalops}) the operator dimensions are $\frac{3}{2}\left(\Delta_{UV}\left(1-\frac{t}{2}\right)-tT_3\right)$ and again we can see many cases of unitarity violations.

We can also ask what happens when we deform by more than one of the $\mu$ operators (i.e. deform by $\Tr \left( \phi_1 \mu_1 +\phi_2\mu_2 + \phi_3\mu_3 \right)$). Using the same reasoning as \cite{Heckman:2010qv} and above we find that the $R$-charge is
\begin{equation}
R_{IR} = \frac{t}{2}R_{\cN=2} + (2-t)I_3 - t(T^{(1)}_3+T^{(2)}_3+T^{(3)}_3),
\end{equation}
where now there is a $T_3$ assigned to each $\phi$ (or equivalently each $\mu$). The value of $t$ is again determined by $a$-maximization to be the same as that given in \eqref{eq:t} with the replacement $k_Gr \rightarrow \sum_i^3 k_G^{(i)}r^{(i)}$. The analysis is qualitatively the same as has been done already.

It is worth noting that these unitarity violations appear to persist even for $k=1$ if one na\"ively uses eq.~\eqref{eq:t}. However, there the theory  consists of free hypermultiplets, and the superpotential deformation $W = \Tr(\phi \mu)$ gives these a mass, so no unitarity problems should occur. In this case, the enhanced symmetry of the free hypers makes $a$-maximization unnecessary, and the theory retains $\cN=2$ SUSY, so eq.~\eqref{eq:t} is not applicable.

We also note that for theories exhibiting unitarity violating operators, this violation could be remedied by an emergent IR symmetry which would require $a$-maximization to be done again, as in \cite{Kutasov:2003iy}. Thus, the apparent violation of unitarity by a few operators does not necessarily mean that the theory is not conformal.
\section{Theories Without Non-Abelian Flavor Symmetries}
\label{sec:wopunc}

In this section we construct an interesting family of new SCFTs using the $T_{N,k}$ theories as building blocks. This approach mirrors that of \cite{Bah:2012dg,Benini:2009mz, Bah:2011vv} which used $T_N$'s coupled by gauging $SU(N)_{diag} \subset SU(N)_1 \times SU(N)_2$, where $SU(N)_1$ and $SU(N)_2$ can belong to different $T_N$'s. The introduction of $T_{N,k}$'s creates significant differences, and the classification of the allowed theories is significantly more complex than that of the $T_N$ quivers. In this section we examine various aspects of these theories and in particular find a puzzle relating to the dimensions of the conformal manifolds.

\subsection{B\texorpdfstring{$^3$}{3}W Theories}
\label{subsec:b3w}

\begin{figure}[ht]\begin{center}
\includegraphics[scale=0.1]{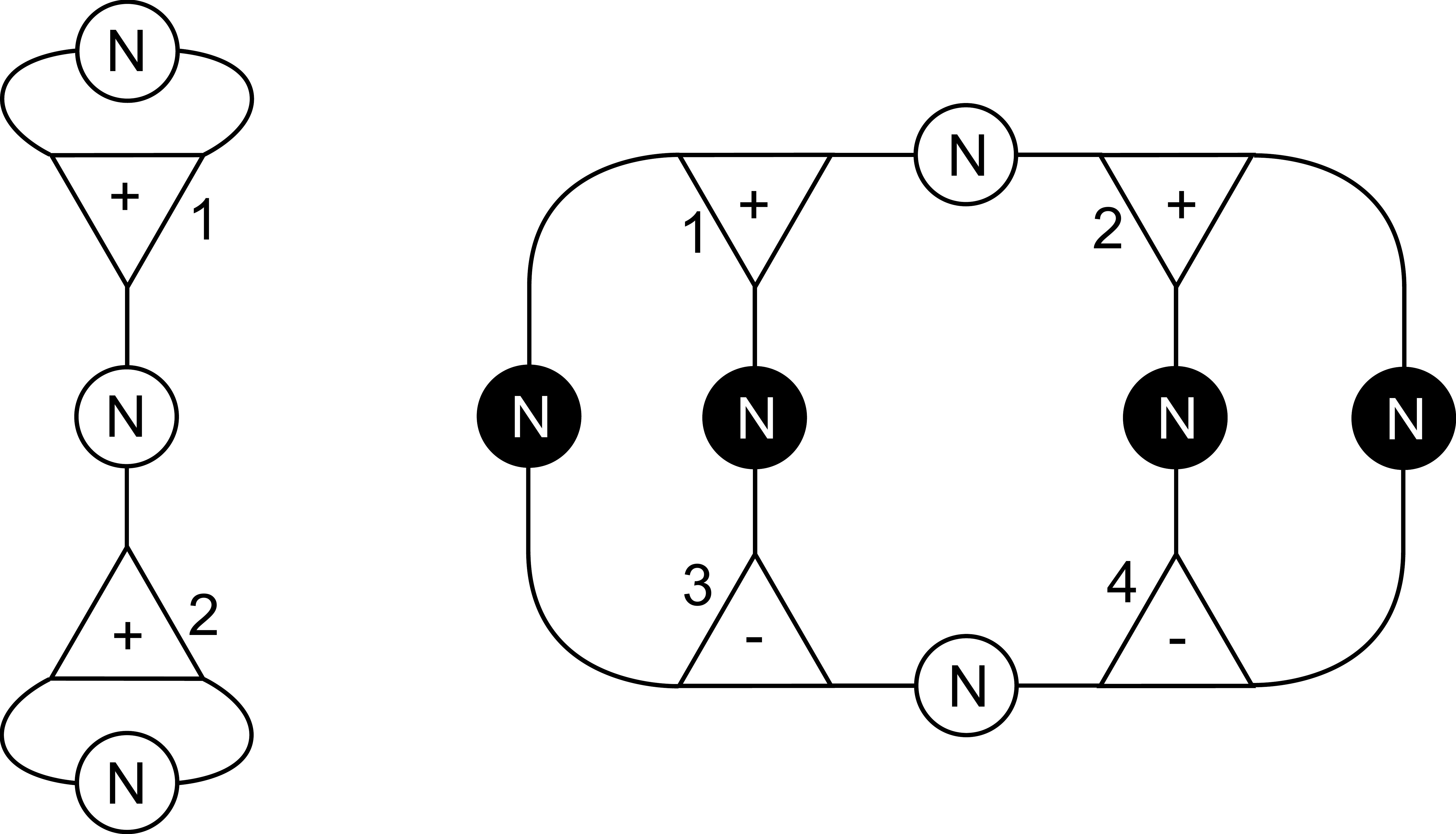}
\caption{Two examples of B$^3$W theories: one with genus two (left) and one with genus three (right).}
\label{fig:b3wex}
\end{center}\end{figure}

We begin with a brief review of the theories in  \cite{Bah:2012dg,Bah:2011vv}, which we refer to as the B$^3$W theories. These theories are constructed by taking a collection of $T_N$ theories and gauging an $SU(N)_{diag} \subset SU(N) \times SU(N)$. The vector multiplet associated to $SU(N)_{diag}$ can be either $\cN=1$ or $\cN=2$. These theories can usefully be pictured by generalized quivers, where we use the convention that white circles are $\cN=2$ multiplets, while black circles are $\cN=1$. In addition to a $U(1)$ $R$-symmetry, these theories also possess exactly one anomaly-free non-R $U(1)$ global symmetry $\cF$. When $\Tr \cF \neq 0$, this can mix with the $R$-symmetry, and $a$-maximization is necessary. A useful choice of $R$-symmetry is
\begin{align}
R_0 = R_{\cN=1} + \frac{1}{6}\sum_i J_i + \frac{1}{3}\sum_A F_A,
\end{align}
and the non-R anomaly-free $U(1)$ symmetry can be taken to be
\begin{equation}
\cF = \sum_i \sigma_iJ_i + 2 \sum_A \sigma_A F_A,
\end{equation}
where $F_A$ is the $U(1)$ symmetry associated to the adjoint chiral superfield living in the $A$-th $\cN=2$ vector multiplet, normalized so that $F_A(\Phi_B)=\delta_{AB}$, and assigning a sign $\sigma_i= \pm 1$ to each $T_N$. The B$^3$W theories follow the rule that $T_N$'s of opposite sign must be connected by a shaded node and $T_N$'s of the same sign must be connected by an unshaded node. One then also assigns a sign $\sigma_A = \pm 1$ to each $\cN=2$ vector multiplet, depending on whether it connects two $T_N$'s of positive or negative sign, respectively. Two examples of these theories are given in figure \ref{fig:b3wex}.

The supergravity duals of these theories were also found in \cite{Bah:2012dg,Bah:2011vv}. In this construction, which generalizes the famous Maldacena-Nu\~nez result \cite{Maldacena:2000mw}, the authors found the near-horizon geometries for an infinite family of $\cN=1$ theories that come from M5-branes wrapping a Riemann surface. The theories are specified by two integer parameters $p$ and $q$, which for $p,q \geq 0$ are  dual to the above quivers. In the UV, this geometry can also be thought of as M5-branes wrapping a Riemann surface $\Sigma_g$ inside a Calabi-Yau, such that the total space is a decomposable line bundle $\cL_1 \oplus \cL_2 \rightarrow \Sigma_g$, with $p$ and $q$ being the Chern numbers of each factor in the bundle. The Calabi-Yau condition then requires $p+q=2g-2$. In the dual field theory, $p$ and $q$  have the interpretation of being the number of $\sigma_i$ of each sign. One check of the duality of the two sides of the AdS/CFT correspondence is the leading-order agreement between the central charges computed on either side; the next-order agreement was found in \cite{Baggio:2014hua}.

Another check of the correspondence is given by the dimensions of the conformal manifolds for these theories, which is $4g-3$. This quantity can be easily computed in the field theory via either Leigh-Strassler \cite{Leigh:1995ep} or the technique of \cite{Green:2010da}. Geometrically, the marginal deformations can be thought of as the $3g-3$ complex structure deformations of the Riemann surface along with the $g$ allowed shifts of the Wilson lines around each cycle by a flat connection, for a total of $4g-3$.

\subsection{Our Setup}
\label{subsec:oursetup}

Here, we will use $T_{N,k}$'s to construct analogs of the B$^3$W theories. This change leads to some profound differences; in particular, we no longer have a known AdS solution that we can use to check our answers. Nevertheless, we will provide some evidence that these constructions lead to interesting new SCFTs in the IR.

\begin{figure}[ht]\begin{center}
\includegraphics[scale=0.15]{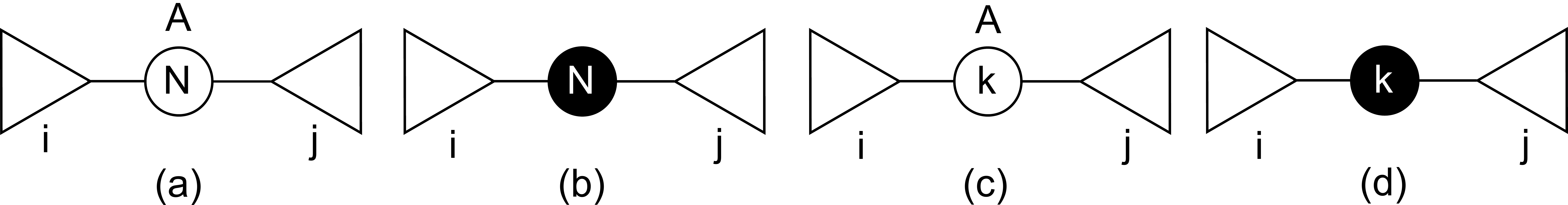}
\caption{The four different ways of coupling $T_{N,k}$'s to vector multiplets. A triangle denotes a $T_{N,k}$, a shaded circle with an $n$ denotes an $\cN=1$ $SU(n)$ vector multiplet and an unshaded circle with an $n$ denotes an $\cN=2$ $SU(n)$ vector multiplet.}
\label{fig:gauge}
\end{center}\end{figure}

The first thing we must determine is how to couple $T_{N,k}$'s to vector multiplets by gauging diagonal subgroups of flavor symmetries. There are four different ways of doing this, as shown in figure \ref{fig:gauge}.
The first two gaugings, a and b, are the same as we had with the B$^3$W theories, and gaugings c and d are new. The first of these new gaugings is an $SU(k)$ gauging using an $\cN=2$ vector multiplet and the second is an $SU(k)$ gauging with an $\cN=1$ vector multiplet. The 1-loop beta function coefficients $b_0$ for the gauge couplings for each of these gauge groups are:
\begin{align}
SU(N) \nonumber \\
&{\rm(a)} \,\, \cN=2 \Rightarrow -b_0=2T(G) - 2 \tfrac{1}{2} k_{SU(N)} = 2N-2N = 0  \nonumber \\
&{\rm(b)}\,\,  \cN=1 \Rightarrow  -b_0=3T(G) - 2 \tfrac{1}{2} k_{SU(N)} = 3N-2N = N  \nonumber \\
SU(k) \nonumber \\
&{\rm(c)} \,\, \cN=2 \Rightarrow  -b_0=2T(G) - 2 \tfrac{1}{2} k_{SU(k)} = 2(k)-2(k+1) = -2  \nonumber \\
&{\rm(d)}\,\, \cN=1 \Rightarrow  -b_0=3T(G) - 2 \tfrac{1}{2} k_{SU(k)} = 3(k)-2(k+1) = k-2 \nonumber
\end{align}
Of the two new gaugings, only the $\cN=1$ $SU(k)$ will become strongly coupled in the IR. The $\cN=2$ gauging goes free in the IR. In the theories we wish to construct we will not be interested in this type of gauging, so we will only consider gaugings a, b, and d. 

The theories we study here will, as before, be constructed by taking an even number of $T_{N,k}$'s and gauging diagonal subgroups until there is no non-Abelian flavor symmetry left. Note that one difference with the B$^3$W theories is that here we only gauge $SU(N)$ or $SU(k)$ inside $SU(N)^2 \times SU(k) \times U(1)$. This means that the theories that we construct will have a residual $U(1)^n$, where $n$ is the number of $T_{N,k}$'s, in addition to any additional anomaly-free $U(1)$ which is a linear combination of $J_i$'s and $F_A$'s. Each of these residual $U(1)$ factors is trace-free, so they will not mix with the IR $R$-symmetry.

These theories have an IR $R$-symmetry of the form
\begin{equation}
R_0 = R_{\cN=1}+\sum_{i}\alpha_iJ_i + \sum_A\beta_AF_A.
\end{equation}
The anomaly-free condition $\Tr \, R_0T^aT^b=0$ gives us constraints on the constants $\alpha_i,\beta_A$, and there are additional constraints from enforcing that superpotential terms $\mu \Phi$, which are necessary for an $\cN=2$ gauging, have $R$-charge two. For gaugings of type a, b and d in figure \ref{fig:gauge} these constraints are as follows:
\begin{align}
SU(N) \text{\ \ } &\cN=2 \Rightarrow \alpha_i=\alpha_j=\frac{1}{2}\beta_A, \nonumber \\
&\cN=1 \Rightarrow \alpha_i+\alpha_j=\frac{1}{3}, \nonumber \\
SU(k) \text{\ \ } &\cN=1 \Rightarrow \alpha_i+\alpha_j=\frac{k}{k+1}-\frac{2}{3}.
\label{eq:Rrules}
\end{align}
Any anomaly-free additional $U(1)$ current will be of the form
\begin{equation}
\cF = \sum_{i=1}\mu_iJ_i + \sum_A\nu_AF_A,
\label{eq:F}
\end{equation}
and the anomaly-free constraint $\Tr \cF T^aT^b=0$, along with the constraint  $\cF(\mu \Phi)=0$, imposes the following:
\begin{align}
\cN=2 & \Rightarrow \mu_i=\mu_j=\frac{1}{2}\nu_A, \nonumber \\
\cN=1 & \Rightarrow \mu_i+\mu_j=0.
\label{eq:Frules}
\end{align}

An interesting difference between the original B$^3$W theories and the theories we wish to construct here is that the additional anomaly-free $U(1)$ symmetry will always be traceless. One way to see this is to note that because each $T_{N,k}$ has only one $SU(k)$ factor, the $T_{N,k}$'s come in pairs connected by $\cN=1$ $SU(k)$ vector multiplets. For a pair that connects the $i^{\rm th}$ and $j^{\rm th}$ $T_{N,k}$, the anomaly-free constraint on $\cF$ is that $\mu_i=-\mu_j$. This means that the first term in \eqref{eq:F} vanishes. One can also show that all $\cN=2$ vector multiplets either: (a) come in pairs with cancelling contributions to $\cF$ (i.e. $\nu_A = \nu_B$), or; (b) have $\nu_A=0$ . This means that for the theories we construct here we will never need to use $a$-maximization to determine the IR $R$-symmetry.

\subsection{A Subclass of Theories}
\label{subsec:subclex1}
\begin{wraptable}{r}{0.45\textwidth}\begin{center}
\begin{tabular}{c c}
Operator & $R$-charge\\
\hline
$\mu_{i}$ & $2-\frac{k}{k+1}$ \\
$u_{n}^{(i)}$ & $n\left(\frac{k}{k+1}\right)$ \\
$\cO_H^{i}$ & $\Delta_{UV}\left(1-\frac{1}{2}\left(\frac{k}{k+1}\right)\right)$\\
$\Phi^n_{A}$ & $n\left(\frac{k}{k+1}\right)$ \\
\end{tabular}
\caption{Some operators of the IR theory with $R$-charges. $\mu_i$ are the $\mu$ operators and $u_n^{(i)}$ ($n \geq 3$) are the Coulomb branch operators for the $i^{\rm th}$ $T_{N,k}$. $\cO_H^i$ are the Higgs branch operators that appear in the SC index (see table \ref{tab:tnklambdalops}) for the $i^{\rm th}$ $T_{N,k}$ and $\Delta_{UV}$ corresponds to the dimension given in table \ref{tab:tnklambdalops}. Finally, $\Phi_A$ are the adjoint chiral superfields belonging to the $A^{\rm th}$ $\cN=2$ vector multiplet.}
\label{tab:subclex1opdims}
\end{center}\end{wraptable}

\begin{figure}[ht]\begin{center}
\includegraphics[scale=0.14]{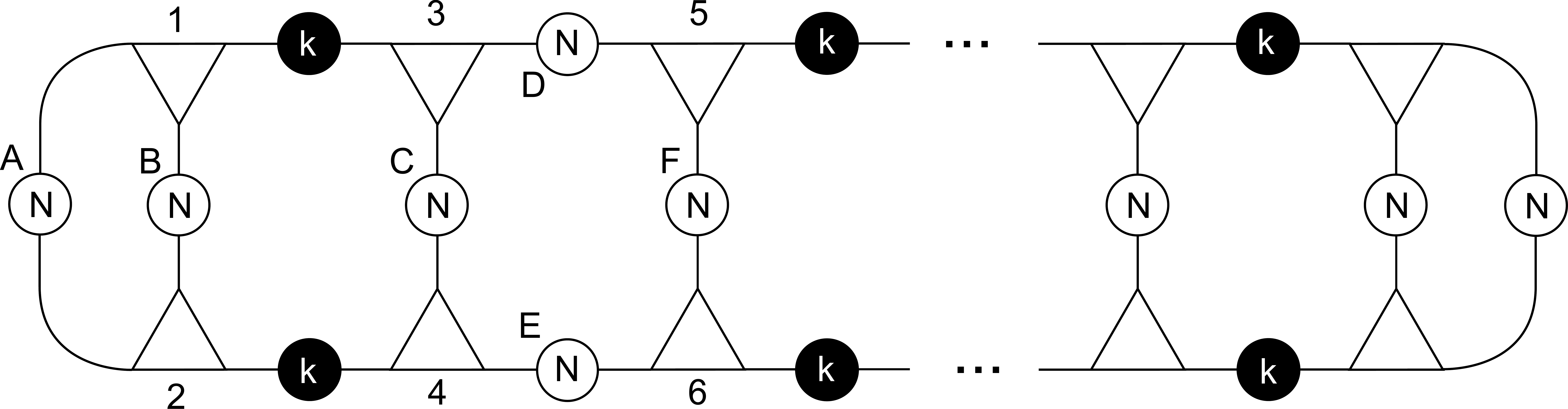}
\caption{The generalized quiver for a theory with no $\cN=1$ $SU(N)$ vector multiplets.}
\label{fig:subclex1}
\end{center}\end{figure}

We first consider a subclass of theories that are constructed in the UV from $T_{N,k}$'s, $\cN=1$ $SU(k)$ vector multiplets, and $\cN=2$ $SU(N)$ vector multiplets. For the moment, we do not include $\cN=1$ $SU(N)$ vectors. An example of one of these theories is given by the quiver in figure \ref{fig:subclex1}. 

Using the rules in \eqref{eq:Rrules} and symmetry of the quiver diagram we find that the IR $R$-symmetry is given by
\begin{align}
R_{IR} =& R_{\cN=1} + \frac{1}{2}\left(\frac{k}{k+1}-\frac{2}{3}\right)\sum_iJ_i  \nonumber \\
&+\left(\frac{k}{k+1}-\frac{2}{3}\right) \sum_AF_A.
\end{align}
The operator dimensions in the IR for this theory are those given in table \ref{tab:subclex1opdims}.

It is easy enough to see that it is impossible to form any gauge-invariant operators that violate the unitarity bound $R \geq \frac{2}{3}$. To determine the dimensions of the conformal manifolds for these theories we use the method of Leigh and Strassler \cite{Leigh:1995ep} (or equivalently \cite{Green:2010da}). From table \ref{tab:subclex1opdims} we see that there are $4g-4$ marginal operators (where $g$ is the genus of the quiver), all of the form $\mu\Phi$. Also, the number of gauge coupling constants is $3g-3$. Finally, there are $4g-5$ constraints coming from fixing anomalous dimensions: $2g-2$ from the $T_{N,k}$'s plus $2g-2$ from the adjoint chiral superfields minus one overall linear combination. This means that the dimension of the conformal manifold for these theories is $3g-2$. At present, we lack a geometric understanding for this number.

\subsection{A Genus Three Example}
\label{subsec:g3ex}
We now look at a particular example where the conformal manifold exhibits a peculiar behavior.
\begin{figure}[ht]\begin{center}
\includegraphics[scale=0.14]{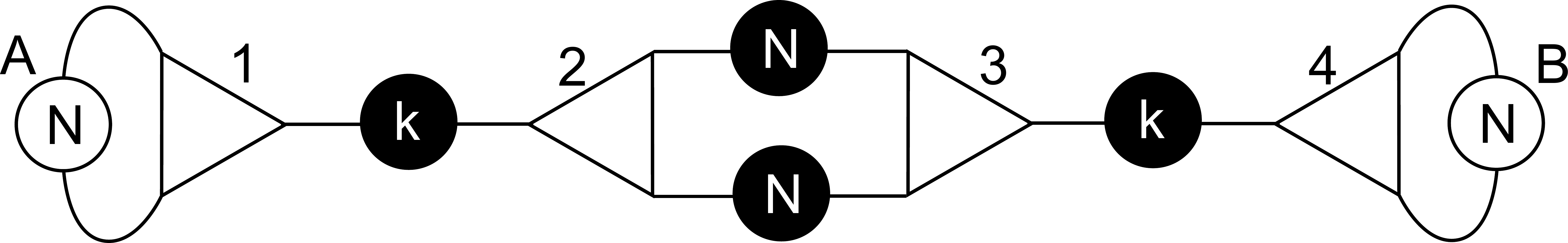}
\caption{A generalized quiver for a theory with genus three. This theory has 4 $T_{N,k}$'s, 2 $\cN=2$ $SU(N)$ vector multiplets, 2 $\cN=1$ $SU(N)$ vector multiplets, and 2 $\cN=1$ $SU(k)$ vector multiplets.}
\label{fig:g3ex}
\end{center}\end{figure}
The theory we wish to consider is given by the quiver diagram in figure \ref{fig:g3ex}. Again, by using the rules in \eqref{eq:Rrules} and symmetry of the quiver we find that the IR $R$ symmetry for this theory is
\begin{equation}
R_{IR} = R_{\cN=1}+\left(\frac{k}{k+1}-\frac{5}{6}\right)\left(J_1+J_4\right)+\frac{1}{6}\left(J_2+J_3\right) +2 \left(\frac{k}{k+1}-\frac{5}{6}\right)\left(F_A+F_B\right).
\label{eq:g3exr0}
\end{equation}
The operators in this theory in the IR and their $R$-charges are shown in table \ref{tab:g3exopdims}.

We once again use the method of Leigh and Strassler \cite{Leigh:1995ep} to determine the dimension of the conformal manifold. We begin by counting the number of marginal operators. From table \ref{tab:g3exopdims} we see that there are 6 operators, $\mu_2^2$, $\mu_3^2$ and $\mu_2\mu_3$ which are all marginal. These come from the 3 $\mu$'s associated to each $T_{N,k}$, related by chiral ring relations \footnote{These relations are $\Tr(\mu_1^2) = \Tr(\mu_2^2)$ for each $T_N$, where $\mu_{1,2}$ transform in the adjoint of the $SU(N)$ factors of the flavor symmetry. See \cite{Benini:2009mz}.}, each transforming in the adjoint of one part of the flavor symmetry. Furthermore, there are 4 marginal operators $\mu_1\Phi_A$, $\mu_4\Phi_B$. There are also 6 gauge coupling constants, which means that the total number of marginal parameters is 16. There are 5 constraints: 4 from fixing the anomalous dimensions of the $T_N$'s,  2  from fixing the anomalous dimensions of the adjoint chiral superfields, minus one overall linear combination. This means that the dimension of the conformal manifold is 11.

\begin{wraptable}{r}{0.35\textwidth}\begin{center}
\begin{tabular}{c c}
Operator & $R$-charge\\
\hline
$\mu_{1,4}$ & $3-2\frac{k}{k+1}$ \\
$\mu_{2,3}$ & $1$ \\
$u_{n}^{(1,4)}$ & $n\left(-1+2\frac{k}{k+1}\right)$ \\
$u_{n}^{(2,3)}$ & $n$ \\
$\cO_H^{1,4}$ & $\Delta_{UV}\left(\frac{3}{2}-\frac{k}{k+1}\right)$ \\
$\cO_H^{2,3}$ & $\Delta_{UV}\left(\frac{1}{2}\right)$ \\
$\Phi^n_{A,B}$ & $n\left(-1+2\frac{k}{k+1}\right)$ \\
\end{tabular}
\caption{The operators of the theory of section \ref{subsec:g3ex} along with their IR $R$-charges.} 
\label{tab:g3exopdims}
\end{center}\end{wraptable}
There is, however, a puzzle. When $k=3$, the number of marginal deformations increase, because the $\Tr\left(\Phi_{A}^4\right)$ and $\Tr\left(\Phi_{B}^4\right)$ operators and the $u_4^{1,4}$ operators (of which there are 4) become marginal. The number of constraints stays the same, and so the dimension of the conformal manifold increases from 11 to 17. This seems a bit strange, since from a geometrical point of view, there is no obvious reason why the $k=3$ theory should be any different from the theories with general $k$. Perhaps this is evidence that these theories are not good SCFTs, but without an AdS dual, it is difficult to say for sure.

\subsection{Another Subclass of Theories}
\label{subsec:subclex2}
\begin{figure}[ht]\begin{center}
\includegraphics[scale=0.15]{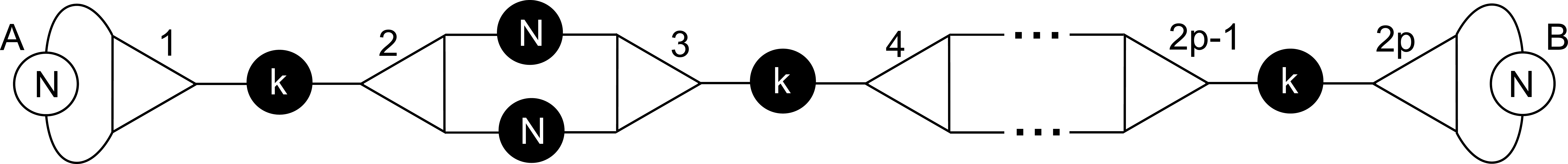}
\caption{The generalized quiver diagram for a subclass of theories. There are $2p=2g-2$ $T_{N,k}s$ where $g$ is the genus of the quiver. Only the two end nodes represent $\cN=2$ gauge groups; the rest are $\cN=1$.}
\label{fig:subclex2}
\end{center}\end{figure}

Inspired by the results of the previous section, we now look at a family of theories that generalize those of the last section. Specifically, we look at theories whose quiver diagrams have the structure given in figure \ref{fig:subclex2}.

The $R$-symmetry for this theory is
\begin{equation}
R_0 = R_{\cN=1} +\sum_{i=1}^{2p}\alpha_iJ_i + \sum_{C=A,B}\beta_CF_C ,
\end{equation}
where
\begin{align}
\alpha_{2n} = \frac{1}{6} -\frac{n-\frac{1}{2}p}{k+1} & & \alpha_{2n-1} = \frac{1}{6} +\frac{n-\frac{1}{2}p-1}{k+1}.
\label{newalphas}
\end{align}
The dimensions of various operators in the theory are given in table \ref{tab:subclex2opdims}. From this table is is easy to construct unitarity violating operators. For instance, we may construct one of these theories with $T_{N,k}$'s and increase the number of $T_{N,k}$'s until $p=k+1$. For this theory the $\Tr (\Phi_{A,B}^m)$ operators have dimension 0. Since the number of problematic operators increases in the large $N$ limit, these theories are most likely not SCFTs.
\begin{wraptable}{r}{0.4\textwidth}\begin{center}
\begin{tabular}{c c}
Operator & $R$-charge \\
\hline
$\mu_{2n}$ & $1+\frac{2n-p}{k+1}$ \\
$\mu_{2n-1}$ & $1-\frac{2n-p-2}{k+1}$ \\
$u_{m}^{(2n)}$ & $m\left(1-\frac{2n-p}{k+1}\right)$ \\
$u_{m}^{(2n-1)}$ & $m\left(1+\frac{2n-p-2}{k+1}\right)$ \\
$\cO_H^{(2n)}$ & $\Delta_{UV}\left(\frac{1}{2}-\frac{n-\frac{1}{2}p}{k+1}\right)$ \\
$\cO_H^{(2n-1)}$ & $\Delta_{UV}\left(\frac{1}{2}-\frac{n-\frac{1}{2}p-1}{k+1}\right)$ \\
$\Phi^m_{A,B}$ & $m\left(1-\frac{p}{k+1}\right)$ \\
\end{tabular}
\caption{The operators of the theory in section \ref{subsec:subclex2} along with their conformal dimensions.} 

\label{tab:subclex2opdims}
\end{center}\end{wraptable}
It is interesting, though disappointing, that these theories (that is, all the theories considered in Section \ref{sec:wopunc}, not just the ones in this subsection), do not fall into a neat classification like their $T_N$ counterparts do. As we have seen, some of the theories we attempted to build out of $T_{N,k}$'s do not appear to be good SCFTs. It would be interesting to find an organizational principle, such as the one in \cite{Bah:2012dg}, that allows us to construct an obvious family of SCFTs. It is of course possible that no such theories are actually conformal, and perhaps the puzzle over the counting of marginal deformations discussed in the previous subsection is evidence of this. To conclusively solve this puzzle once and for all, we would need a method for constructing the AdS duals to these theories, and such an understanding is still lacking. Until then, we will have to regard the results of the present work as merely preliminary.

\section{Conclusions}
\label{sec:outro}
%

In this work, we investigated various properties of an interesting infinite family of theories of class $\cS$, which we call the $T_{N,k}$ theories. These theories generalize Gaiotto's $T_N$ theory and naturally arise when considering various S-duality frames of curves with two maximal punctures and multiple minimal punctures. Using techniques from duality as well as the superconformal index we described various properties of these theories, such as their global anomalies, central charges, and various operators and dimensions. We then used these theories as building blocks for constructing new $\cN=1$ SCFTs, and checked whether the various theories under consideration appeared to be good conformal theories.

Our work raises some interesting questions. The most pressing is, of course, whether or not there exist AdS duals to these theories. Even in the case of the earlier work \cite{Bah:2013aha}, it still remains unclear whether or not AdS duals for $S_\ell$ theories exist, and our understanding of the $\cN=1$ AdS duals with punctured surfaces remains very incomplete. In order to establish beyond a reasonable doubt the existence of the SCFTs in this work as well as \cite{Bah:2013aha}, it remains a pressing problem to find such duals. This would presumably also help us understand the dimension of the conformal manifold, a quantity for which at present we lack a geometric understanding for these theories.

Another interesting question is whether or not there exists a general principle, like the ones found for the B${}^3$W theories, we could use for building analogous quivers out of the $T_{N,k}$'s. We were unable to find such a general principle, though it is possible that one exists. It is also possible that the absence of such a principle, as well as a seeming mismatch between various quantities of interest ({\it e.g.}, the dimension of the conformal manifold) in what would naively be considered different duality frames, may indicate that these theories are not indeed good SCFTs. On the other hand, it is also possible that there is an interesting geometric reason why the $\cN=1$ theories we consider here do not allow the full range of dualities found in the analogous $\cN=2$ cases, and in the absence of a good geometric understanding of these constructions, it may indeed be the most likely possibility that no such dualities exist. This possibility is especially tantalizing, since understanding the geometric origin of such an obstruction would no doubt be of great interest.

The larger question explored by this work is which $\cN=1$ SCFTs can be built out of class $\cS$ building blocks. As we know from our study of general $\cN=1$ theories with weakly coupled matter, it is no easy task to determine when a theory reaches a conformal fixed point in the IR. However, it is not outside the realm of possibility that, by using the techniques employed in this work as well as others, we could find large new tracts of the landscape of $\cN=1$ SCFTs.

%

\begin{center}
\bf{Acknowledgements}
\end{center}
\medskip

We would like to thank David Garner, Edward Hughes, Brenda Penante, Sanjaye Ramgoolam, Felix Rudolph, and Yuji Tachikawa for useful discussions. This work is supported in part by the STFC Standard Grant ST/J000469/1 ``String Theory, Gauge Theory and Duality." JMG is supported by a Queen Mary University of London studentship.

\vspace*{0.2in}

\appendix

\section{Appendix: Extra Results For \texorpdfstring{$\cN=1$}{N=1} Theories}
\label{app:extres}
\renewcommand{\theequation}{A.\arabic{equation}}
\setcounter{equation}{0}

\subsection{\texorpdfstring{$S_{\ell}$}{Sl} Theories With A \texorpdfstring{$T_{N,k_1}$}{TNk1} And A \texorpdfstring{$T_{N,k_2}$}{TNk2}}
\label{app:sltnk1tnk2}

In section \ref{subsec:sltnk} we looked at the $S_{\ell}$ theory with two $T_{N,k}$'s. Here we summarize the results for the $S_{\ell}$ theory with a $T_{N,k_1}$ at one end of the quiver and a $T_{N,k_2}$ at the other end.

As for the $S_{\ell}$ theory with two $T_{N,k}$'s we have an $R$-symmetry which is the same as that given in equation \eqref{eq:r0} and also an additional anomaly-free $U(1)$ symmetry \eqref{eq:addu1}. In contrast to the $S_{\ell}$ theory with two $T_{N,k}$'s it is no longer the case that $\Tr \cF = 0$ when $\ell$ is even and so we must use $a$-maximization for all $\ell$. If we do this we find that the value of $\alpha$ that maximizes $a$ is
\begin{equation}
\widehat{\alpha}=\frac{A+ \sqrt{B}}{C},
\end{equation}
where
\begin{align}
A=&-3N^2 \left(k_1+k_2+\ell\right)+k_1^3+k_2^3-k_1-k_2, \nonumber \\
B=&N^4 \left(18 k_1 \ell +18 k_2 \ell +36 k_1^2+36 k_2^2-102 k_1+72 k_1 k_2-102 k_2+9 \ell ^2+91\right) \nonumber \\
&+N^2 \left(-6 k_1^3 \ell +6 k_1 \ell -6 k_2^3 \ell +6 k_2 \ell -24 k_1^4-24 k_2 k_1^3+10 k_1^3-24 k_2 k_1^2+40 k_1^2 \right. \nonumber \\
& \left. \phantom{+N^2 \ \ } -24 k_2^3 k_1-24 k_2^2 k_1+102 k_1-24 k_2^4+10 k_2^3+40 k_2^2+102 k_2-160\right) \nonumber \\
&+4 k_1^6+8 k_1^5+4 k_1^4-32 k_1^3-32 k_1^2+8 k_2^3 k_1^3+8 k_2^2 k_1^3+8 k_2^3 k_1^2+8 k_2^2 k_1^2 \nonumber \\
&+4 k_2^6+8 k_2^5+4 k_2^4-32 k_2^3-32 k_2^2+64 \nonumber \\
C=&\left(-18 k_1-18 k_2+42\right) N^2+6 k_1^3+12 k_1^2+6 k_1+6 k_2^3+12 k_2^2+6 k_2-48. \nonumber
\end{align}
We do not plot this here as the plots are much the same as those in figure \ref{fig:sltnkalpha} however we note that it approaches
\begin{equation}
\frac{-3 \left(k_1+k_2+\ell\right) +\sqrt{18 \left(k_1+k_2\right) \ell +72 k_1 k_2-102 \left(k_1+k_2\right)+36 \left(k_1^2+k_2^2\right)+9 \ell ^2+91}}{6 \left(-3 k_1-3 k_2+7\right)} \nonumber
\end{equation}
at large $N$. One can verify that this never goes below $-\frac{1}{6}$ and so any gauge-invariant operators that can be constructed satisfy the unitarity bound $R \geq \frac{2}{3}$.

\subsection{\texorpdfstring{$S_{\ell}$}{Sl} Theories With Adjoint Matter}
\label{app:sltnkadj}

In section \ref{subsec:higgs} we looked at what happens if we take the $S_{\ell}$ theory and give a vev to the $k$-th hypermultiplet. The theory that we get in IR is that represented by the bottom quiver diagram in figure \ref{fig:flow}. It is an $S_{\ell}$ theory with an adjoint chiral superfield and extra $Q\Phi\widetilde{Q}$ superpotential terms. For this theory with two $T_{N,k}$'s at each end of the quiver the $R$-symmetry is $R_{IR} = R_0 + \widehat{\alpha} \cF$, where $R_0$ is given in equation \eqref{eq:r0} with $R_0(\Phi)=1$, $\cF$ is given in equation \eqref{eq:fhiggs} and $\widehat{\alpha}$ is found using $a$-maximization to be
\begin{equation}
\widehat{\alpha}=\frac{A+ \sqrt{B}}{C},
\end{equation}
where
\begin{align}
A=& 96 N^3 + 72 \ell N^2-96 N, \nonumber \\
B=& N^6\left(13824 (-1)^l+23040\right) \nonumber \\
&+N^5 \left(20736 (-1)^{k+l}+20736 (-1)^k+13824 l-57600 (-1)^l-57600\right) \nonumber \\
&+N^4 \left(-43200 (-1)^{k+l}-43200 (-1)^k+5184 l^2+44640 (-1)^l+41760\right) \nonumber \\
&+N^3 \left(-25344 (-1)^{k+l}-25344 (-1)^k-13824 l+62208 (-1)^l+62208\right) \nonumber \\
&+N^2 \left(59904 (-1)^{k+l}+59904 (-1)^k-64512(-1)^l-82944\right) \nonumber \\
&+N \left(9216 (-1)^{k+l}+9216 (-1)^k-18432 (-1)^l-18432\right) \nonumber \\
&-18432 (-1)^{k+l}-18432 (-1)^k+18432 (-1)^l+27648, \nonumber \\
C=&\left(288 (-1)^l+288\right) N^3+N^2 \left(432 (-1)^k-792 (-1)^l-792\right) \nonumber \\
&-576 (-1)^k+576 (-1)^l+576. \nonumber
\end{align}

When we do this for the same theory but with a $T_{N,k_1}$ at one end of the quiver and a $T_{N,k_2}$ at the other end, the IR $R$-symmetry is again given by $R_{IR} = R_0 + \widehat{\alpha} \cF$. As the expression for $\widehat{\alpha}$ would take up too much space,  we do not include it here.

\bibliographystyle{JHEP}
\bibliography{tnflows}

\providecommand{\href}[2]{#2}\begingroup\raggedright\begin{thebibliography}{10}

\bibitem{Maldacena:2000mw}
J.~M. Maldacena and C.~Nunez, {\it {Supergravity description of field theories
  on curved manifolds and a no go theorem}},  {\em Int.J.Mod.Phys.} {\bf A16}
  (2001) 822--855, [\href{http://arxiv.org/abs/hep-th/0007018}{{\tt
  hep-th/0007018}}].

\bibitem{Chacaltana:2010ks}
O.~Chacaltana and J.~Distler, {\it {Tinkertoys for Gaiotto Duality}},  {\em
  JHEP} {\bf 1011} (2010) 099, [\href{http://arxiv.org/abs/1008.5203}{{\tt
  arXiv:1008.5203}}].

\bibitem{Gaiotto:2009we}
D.~Gaiotto, {\it {N=2 dualities}},  {\em JHEP} {\bf 1208} (2012) 034,
  [\href{http://arxiv.org/abs/0904.2715}{{\tt arXiv:0904.2715}}].

\bibitem{Romelsberger:2005eg}
C.~Romelsberger, {\it {Counting chiral primaries in N = 1, d=4 superconformal
  field theories}},  {\em Nucl.Phys.} {\bf B747} (2006) 329--353,
  [\href{http://arxiv.org/abs/hep-th/0510060}{{\tt hep-th/0510060}}].

\bibitem{Kinney:2005ej}
J.~Kinney, J.~M. Maldacena, S.~Minwalla, and S.~Raju, {\it {An Index for 4
  dimensional super conformal theories}},  {\em Commun.Math.Phys.} {\bf 275}
  (2007) 209--254, [\href{http://arxiv.org/abs/hep-th/0510251}{{\tt
  hep-th/0510251}}].

\bibitem{Gadde:2011ik}
A.~Gadde, L.~Rastelli, S.~S. Razamat, and W.~Yan, {\it {The 4d Superconformal
  Index from q-deformed 2d Yang-Mills}},  {\em Phys.Rev.Lett.} {\bf 106} (2011)
  241602, [\href{http://arxiv.org/abs/1104.3850}{{\tt arXiv:1104.3850}}].

\bibitem{Bah:2012dg}
I.~Bah, C.~Beem, N.~Bobev, and B.~Wecht, {\it {Four-Dimensional SCFTs from
  M5-Branes}},  {\em JHEP} {\bf 1206} (2012) 005,
  [\href{http://arxiv.org/abs/1203.0303}{{\tt arXiv:1203.0303}}].

\bibitem{Bah:2013qya}
I.~Bah, {\it {Quarter-BPS $AdS_{5}$ solutions in M-theory with a $T^{2}$ bundle
  over a Riemann surface}},  {\em JHEP} {\bf 1308} (2013) 137,
  [\href{http://arxiv.org/abs/1304.4954}{{\tt arXiv:1304.4954}}].

\bibitem{Xie:2013gma}
D.~Xie, {\it {M5 brane and four dimensional N = 1 theories I}},  {\em JHEP}
  {\bf 1404} (2014) 154, [\href{http://arxiv.org/abs/1307.5877}{{\tt
  arXiv:1307.5877}}].

\bibitem{Xie:2013rsa}
D.~Xie and K.~Yonekura, {\it {Generalized Hitchin system, Spectral curve and
  $\mathcal{N} = 1$ dynamics}},  {\em JHEP} {\bf 1401} (2014) 001,
  [\href{http://arxiv.org/abs/1310.0467}{{\tt arXiv:1310.0467}}].

\bibitem{Bah:2013aha}
I.~Bah and N.~Bobev, {\it {Linear Quivers and N=1 SCFTs from M5-branes}},
  \href{http://arxiv.org/abs/1307.7104}{{\tt arXiv:1307.7104}}.

\bibitem{Bonelli:2013pva}
G.~Bonelli, S.~Giacomelli, K.~Maruyoshi, and A.~Tanzini, {\it $\mathcal{N}=1$
  geometries via m-theory},  {\em JHEP} {\bf 1310} (2013) 227,
  [\href{http://arxiv.org/abs/1307.7703}{{\tt arXiv:1307.7703}}].

\bibitem{Benini:2009mz}
F.~Benini, Y.~Tachikawa, and B.~Wecht, {\it {Sicilian gauge theories and N=1
  dualities}},  {\em JHEP} {\bf 1001} (2010) 088,
  [\href{http://arxiv.org/abs/0909.1327}{{\tt arXiv:0909.1327}}].

\bibitem{Bah:2011je}
I.~Bah and B.~Wecht, {\it {New N=1 Superconformal Field Theories In Four
  Dimensions}},  {\em JHEP} {\bf 1307} (2013) 107,
  [\href{http://arxiv.org/abs/1111.3402}{{\tt arXiv:1111.3402}}].

\bibitem{Bah:2011vv}
I.~Bah, C.~Beem, N.~Bobev, and B.~Wecht, {\it {AdS/CFT Dual Pairs from
  M5-Branes on Riemann Surfaces}},  {\em Phys.Rev.} {\bf D85} (2012) 121901,
  [\href{http://arxiv.org/abs/1112.5487}{{\tt arXiv:1112.5487}}].

\bibitem{Maruyoshi:2013hja}
K.~Maruyoshi, Y.~Tachikawa, W.~Yan, and K.~Yonekura, {\it {N=1 dynamics with
  $T_{N}$ theory}},  {\em JHEP} {\bf 1310} (2013) 010,
  [\href{http://arxiv.org/abs/1305.5250}{{\tt arXiv:1305.5250}}].

\bibitem{Agarwal:2013uga}
P.~Agarwal and J.~Song, {\it {New N=1 Dualities from M5-branes and
  Outer-automorphism Twists}},  {\em JHEP} {\bf 1403} (2014) 133,
  [\href{http://arxiv.org/abs/1311.2945}{{\tt arXiv:1311.2945}}].

\bibitem{Tachikawa:2011ea}
Y.~Tachikawa and K.~Yonekura, {\it {N=1 curves for trifundamentals}},  {\em
  JHEP} {\bf 1107} (2011) 025, [\href{http://arxiv.org/abs/1105.3215}{{\tt
  arXiv:1105.3215}}].

\bibitem{Agarwal:2014rua}
P.~Agarwal, I.~Bah, K.~Maruyoshi, and J.~Song, {\it {Quiver Tails and N=1 SCFTs
  from M5-branes}},  \href{http://arxiv.org/abs/1409.1908}{{\tt
  arXiv:1409.1908}}.

\bibitem{Giacomelli:2014rna}
S.~Giacomelli, {\it {Four dimensional superconformal theories from M5 branes}},
   \href{http://arxiv.org/abs/1409.3077}{{\tt arXiv:1409.3077}}.

\bibitem{Chacaltana:2011ze}
O.~Chacaltana and J.~Distler, {\it {Tinkertoys for the $D_N$ series}},
  \href{http://arxiv.org/abs/1106.5410}{{\tt arXiv:1106.5410}}.

\bibitem{Chacaltana:2012ch}
O.~Chacaltana, J.~Distler, and Y.~Tachikawa, {\it {Gaiotto Duality for the
  Twisted $A_{2N-1}$ Series}},  \href{http://arxiv.org/abs/1212.3952}{{\tt
  arXiv:1212.3952}}.

\bibitem{Chacaltana:2012zy}
O.~Chacaltana, J.~Distler, and Y.~Tachikawa, {\it {Nilpotent orbits and
  codimension-two defects of 6d $\mathcal{N}=(2,0)$ theories}},  {\em
  Int.J.Mod.Phys.} {\bf A28} (2013) 1340006,
  [\href{http://arxiv.org/abs/1203.2930}{{\tt arXiv:1203.2930}}].

\bibitem{Chacaltana:2013oka}
O.~Chacaltana, J.~Distler, and A.~Trimm, {\it {Tinkertoys for the Twisted
  D-Series}},  \href{http://arxiv.org/abs/1309.2299}{{\tt arXiv:1309.2299}}.

\bibitem{Chacaltana:2014jba}
O.~Chacaltana, J.~Distler, and A.~Trimm, {\it {Tinkertoys for the $E_6$
  Theory}},  \href{http://arxiv.org/abs/1403.4604}{{\tt arXiv:1403.4604}}.

\bibitem{Chacaltana:2015bna}
O.~Chacaltana, J.~Distler, and A.~Trimm, {\it {Tinkertoys for the Twisted $E_6$
  Theory}},  \href{http://arxiv.org/abs/1501.0035}{{\tt arXiv:1501.0035}}.

\bibitem{Nanopoulos:2010zb}
D.~Nanopoulos and D.~Xie, {\it {Hitchin Equation, Irregular Singularity, and
  $N=2$ Asymptotical Free Theories}},
  \href{http://arxiv.org/abs/1005.1350}{{\tt arXiv:1005.1350}}.

\bibitem{Gaiotto:2011xs}
D.~Gaiotto, G.~W. Moore, and Y.~Tachikawa, {\it {On 6d N=(2,0) theory
  compactified on a Riemann surface with finite area}},
  \href{http://arxiv.org/abs/1110.2657}{{\tt arXiv:1110.2657}}.

\bibitem{Minahan:1996fg}
J.~A. Minahan and D.~Nemeschansky, {\it {An N=2 superconformal fixed point with
  E(6) global symmetry}},  {\em Nucl.Phys.} {\bf B482} (1996) 142--152,
  [\href{http://arxiv.org/abs/hep-th/9608047}{{\tt hep-th/9608047}}].

\bibitem{Argyres:2007cn}
P.~C. Argyres and N.~Seiberg, {\it {S-duality in N=2 supersymmetric gauge
  theories}},  {\em JHEP} {\bf 0712} (2007) 088,
  [\href{http://arxiv.org/abs/0711.0054}{{\tt arXiv:0711.0054}}].

\bibitem{Argyres:2007tq}
P.~C. Argyres and J.~R. Wittig, {\it {Infinite coupling duals of N=2 gauge
  theories and new rank 1 superconformal field theories}},  {\em JHEP} {\bf
  0801} (2008) 074, [\href{http://arxiv.org/abs/0712.2028}{{\tt
  arXiv:0712.2028}}].

\bibitem{Gaiotto:2009gz}
D.~Gaiotto and J.~Maldacena, {\it {The Gravity duals of N=2 superconformal
  field theories}},  \href{http://arxiv.org/abs/0904.4466}{{\tt
  arXiv:0904.4466}}.

\bibitem{Gadde:2010te}
A.~Gadde, L.~Rastelli, S.~S. Razamat, and W.~Yan, {\it {The Superconformal
  Index of the $E_6$ SCFT}},  {\em JHEP} {\bf 1008} (2010) 107,
  [\href{http://arxiv.org/abs/1003.4244}{{\tt arXiv:1003.4244}}].

\bibitem{Intriligator:2003jj}
K.~A. Intriligator and B.~Wecht, {\it {The Exact superconformal R symmetry
  maximizes a}},  {\em Nucl.Phys.} {\bf B667} (2003) 183--200,
  [\href{http://arxiv.org/abs/hep-th/0304128}{{\tt hep-th/0304128}}].

\bibitem{Heckman:2010qv}
J.~J. Heckman, Y.~Tachikawa, C.~Vafa, and B.~Wecht, {\it {N = 1 SCFTs from
  Brane Monodromy}},  {\em JHEP} {\bf 1011} (2010) 132,
  [\href{http://arxiv.org/abs/1009.0017}{{\tt arXiv:1009.0017}}].

\bibitem{Kutasov:2003iy}
D.~Kutasov, A.~Parnachev, and D.~A. Sahakyan, {\it {Central charges and
  $U(1)_R$ symmetries in N=1 superYang-Mills}},  {\em JHEP} {\bf 0311} (2003)
  013, [\href{http://arxiv.org/abs/hep-th/0308071}{{\tt hep-th/0308071}}].

\bibitem{Baggio:2014hua}
M.~Baggio, N.~Halmagyi, D.~R. Mayerson, D.~Robbins, and B.~Wecht, {\it {Higher
  Derivative Corrections and Central Charges from Wrapped M5-branes}},
  \href{http://arxiv.org/abs/1408.2538}{{\tt arXiv:1408.2538}}.

\bibitem{Leigh:1995ep}
R.~G. Leigh and M.~J. Strassler, {\it {Exactly marginal operators and duality
  in four-dimensional N=1 supersymmetric gauge theory}},  {\em Nucl.Phys.} {\bf
  B447} (1995) 95--136, [\href{http://arxiv.org/abs/hep-th/9503121}{{\tt
  hep-th/9503121}}].

\bibitem{Green:2010da}
D.~Green, Z.~Komargodski, N.~Seiberg, Y.~Tachikawa, and B.~Wecht, {\it {Exactly
  Marginal Deformations and Global Symmetries}},  {\em JHEP} {\bf 1006} (2010)
  106, [\href{http://arxiv.org/abs/1005.3546}{{\tt arXiv:1005.3546}}].

\end{thebibliography}\endgroup

\end{document}